\documentclass{aa}  
\usepackage{booktabs}
\usepackage{placeins}
\usepackage{graphicx}
\usepackage{txfonts}
\usepackage{xcolor}
\usepackage{lipsum}
\usepackage{adjustbox}
\usepackage{soul}
\usepackage{float}
\usepackage{multirow}
\usepackage{rotating} 
\usepackage[markup=underlined]{changes}
\usepackage{hyperref}

\hypersetup{colorlinks, urlcolor=blue, citecolor=blue, linkcolor=black}

\begin{document} 

   \title{The system initial mass function of Orion OB1b from probabilistic Gaia DR3 memberships}


   \titlerunning{The IMF of Orion OB1b}

    \author{Rodrigo Cabral\inst{1} \corrauth{rcabral@fcien.edu.uy}\and
    Juan José Downes \inst{1} \email{jdownes@fcien.edu.uy}\and
    Rafael Bertolotto-Stefanelli \inst{2} \email{rbertolotto@arizona.edu} \and
    Genaro Suárez\inst{3} \email{gsuarez@amnh.org}}
            
   \institute{Departamento de Astronom\'ia, Facultad de Ciencias, Universidad de la Rep\'ublica, Igu\'a 4225, Montevideo, CP 11400, Uruguay\\
         \and
             Steward Observatory, University of Arizona, 933 North Cherry Avenue, Tucson, AZ 85721, USA\\
         \and American Museum of Natural History, 200 Central Park W, New York, NY 10024, USA
             }

   \date{Received 4 June 2026; accepted 27 November 2026}

 
\abstract
{Whether the initial mass function (IMF) varies with environment remains a central question in star formation. The multiple kinematically distinct populations of Orion OB1b provide an ideal laboratory to test possible IMF variations.}
{We aim to characterise the stellar populations of Orion OB1b by identifying its constituent groups, assigning membership probabilities to individual sources, and determining their IMFs.}
{Using Gaia DR3 astrometry and photometry, we identified stellar groups through clustering analysis and assigned membership probabilities with a probabilistic mixture model accounting for astrometric uncertainties and field contamination. Stellar masses were derived from extinction-corrected photometry using pre-main-sequence evolutionary models.}
{We identified eight kinematically distinct groups and a non-clustered population, deriving ages and IMFs spanning ${\sim}0.08$--$10\,M_\odot$ for six groups, all consistent with being gravitationally unbound. Each IMF was fitted with a lognormal distribution below $1M_\odot$ and a power law above. The high-mass regime is statistically indistinguishable across all groups, while four groups also exhibit consistent IMFs over the full mass range. The remaining two groups show differences that may reflect membership uncertainties rather than intrinsic IMF variations. The combined IMF is characterised by $m_c=0.27^{+0.02}_{-0.02},M_\odot$, $\sigma=0.45^{+0.03}_{-0.03}$, and $\Gamma=1.85^{+0.16}_{-0.14}$. The non-clustered population exhibits a broader and more uncertain IMF, although residual field contamination cannot be ruled out.} 
{Our results support previous evidence 
for the complex internal structure of 
OB1b. The IMF 
in OB1b appears broadly consistent across its kinematically distinct groups and compatible with other young populations in Orion. Although some components show apparent deviations, current data cannot establish whether these are intrinsic or arise from membership uncertainties in the sparsest populations. Deeper photometric and spectroscopic follow-up will be required to clarify this.}


   \maketitle

\nolinenumbers
%

\section{Introduction}

The initial mass function (IMF) describes the distribution of stellar masses at the time of formation of a stellar population. It is a fundamental outcome of the star formation process and a key input for many astrophysical models \citep[e.g.][]{hopkins2018}. Although the dependence of its shape on environmental properties remains debated \citep[e.g.,][]{offner_origin_2014}, the IMF is expected to provide insights into the processes and conditions involved in star formation across different stellar mass ranges. These factors include the turbulence and metallicity of the progenitor gas, as well as the presence of nearby massive stars \citep[e.g.,][]{hennebelle2024}. In particular, the shape of the IMF may offer clues about the mechanisms that regulate the formation of very low-mass stars and brown dwarfs \citep[e.g.,][]{whitworth_formation_2007}, as well as those operating in the high-mass regime \citep[e.g.,][]{bastian_universal_2010}, both of which are still not fully understood.

Since the seminal work of \cite{salpeter_luminosity_1955}, numerous IMFs have been derived observationally in the solar neighbourhood, both for the field \citep[e.g.,][]{chabrier_galactic_2003,sollima_stellar_2019,kirkpatrick2024} 
and for young clusters and associations 
\citep[e.g.,][]{da_rio_initial_2012,pena_ramirez_new_2012,drass_bimodal_2016,suarez_system_2019, bertolotto2026}. Despite these efforts, additional IMF determinations covering a broad range of stellar masses within each population and spanning a variety of environmental conditions are still needed.

Particularly valuable are IMF determinations for populations belonging to the same star-forming complex, which likely share properties such as metallicity, while differing in quantities such as total mass, spatial density, and age, potentially affecting both the dynamical evolution of the groups and the radiative feedback from massive stars. Observationally, these populations have the advantage that, given their young ages, dynamical and stellar evolutionary effects are expected to have a limited impact on the observed mass distribution. This reduces the need for corrections accounting for dynamical evolution, stellar evolution, and the star formation history, which are required for older populations. As a result, IMF comparisons become more meaningful for understanding the underlying processes of star formation.

Among nearby star-forming regions, Orion is one of the most important laboratories for IMF studies because it contains multiple young stellar populations spanning a wide range of masses, stellar densities, and early evolutionary stages.
We present the first determination of the IMF in the Orion OB1b sub-association based on Gaia DR3 \citep{gaia_collaboration_gaia_2022} data and compare the results with previous determinations for other populations within the Orion star-forming region. In Section \ref{sec:OrionOB1b}, we summarise the current view of Orion OB1b and the stellar groups identified within it. The selection of photometric and kinematic candidates for young stars and brown dwarfs is presented in Section \ref{sec:candidates}. The statistical determination of membership probabilities for the identified stellar groups is discussed in Section \ref{sec:memberships}. The luminosity and mass functions are presented in Section \ref{sec:LandMfunctions} and discussed further in Section \ref{sec:discussion}. Finally, Section \ref{sec:conclusions} summarises our results and conclusions.

\section{Present view of Orion OB1b}\label{sec:OrionOB1b}

\subsection{Boundaries of OB1b in the Orion complex}
\label{sec:OB1bGral}

The Orion OB1 star-forming region is one of the nearest ($d\sim400~$pc) and largest OB associations \citep[see][for an extensive review]{briceno_dispersed_2008}. Its proximity, favourable direction towards the Galactic anticentre, rich substructure, and relatively low extinction ($A_V\lesssim1$~mag) make it an important laboratory for studies of young stellar populations 
and the IMF.

\cite{blaauw_o_1964} divided Orion OB1 into four sub-regions: OB1a, OB1b, OB1c, and OB1d, identifying OB1a as the oldest and OB1d as the youngest. This classification reflected the idea of sequential star formation in Orion and was widely adopted in subsequent studies. More detailed boundaries of OB1b were established by \cite{warren_photometric_1977}, who defined it as an approximately rectangular region around the Orion Belt stars and subdivided it into three regions associated with each of them. This subdivision was based on a distance gradient then believed to exist among the Belt stars, but was later revised by \cite{guetter_spectroscopic_1981} and \cite{brown_orion_1994}, who found no significant differences among those populations.

\begin{figure}
    \centering
    \includegraphics[width=\columnwidth]{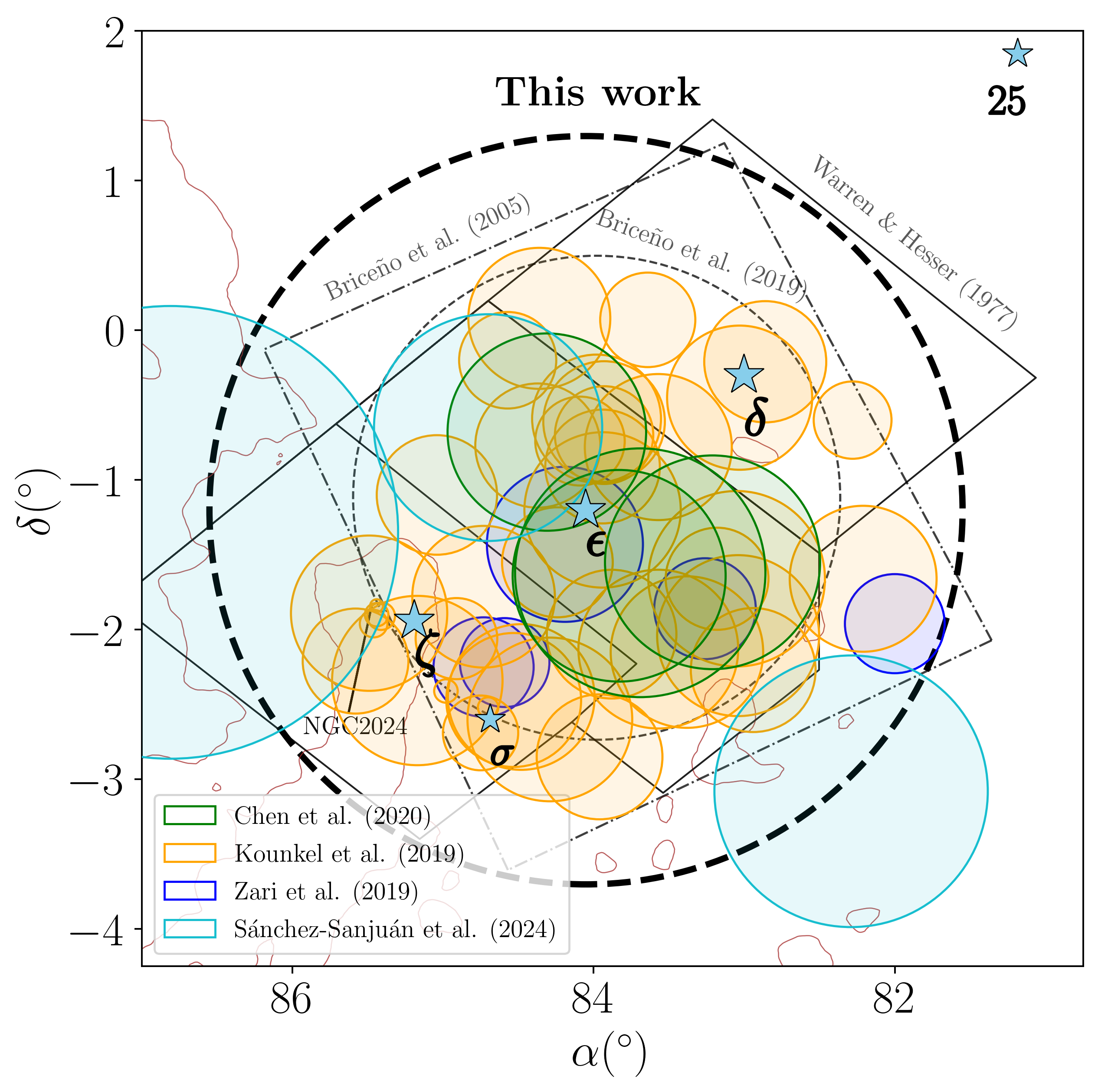}
    \caption{Projected spatial distribution of stellar groups within Orion OB1b. 
    Thick dashed black circle: region analysed in this work. Labeled thin black lines: literature boundaries. 
    Green, orange, blue and cyan circles indicate groups from \cite{chen2020}, \cite{kounkel_apogee-2_2018}, \cite{zari2019}, and \cite{sanchez2024}, respectively. Circle radii correspond to the average of the $\alpha$ and $\delta$ standard deviations reported by the authors. Brown contours show regions with $A_V > 1$~mag 
    from \cite{schlegel1998}.}
    \label{figOB1boverview}
\end{figure}

More recent studies have adopted broader and more general boundaries for OB1b on the basis of spectroscopically confirmed young low-mass stars \citep[e.g.,][]{briceno_cida_2005,briceno_dispersed_2008,briceno_cida_2019}. 
\autoref{figOB1boverview} shows a compilation of the boundaries adopted in the literature, as well as the region analysed in this work, which provides complete coverage of the currently known extent of OB1b.

\subsection{Orion in the eras of spectroscopic follow-ups and Gaia}
\label{sec:neweras}

The presence of overlapping kinematically distinct populations towards OB1b was already recognised before the Gaia era. One of the best studied stellar populations in the region is the $\sigma$ Ori cluster located in the southeastern part of OB1b \citep[e.g.,][]{caballero_brightest_2007,hernandez2007, jeffries_kinematic_2006, pena_ramirez_new_2012,damian2023}. \cite{jeffries_kinematic_2006} found a bimodal radial-velocity distribution within the cluster region, suggesting contamination from another kinematically distinct young population. From a sample of photometrically selected and spectroscopically confirmed young candidates, \cite{caballero_young_2008} identified two populations around $\epsilon$ and $\delta$ Ori, which they argued could be more massive and more spatially extended than $\sigma$ Ori itself. This scenario was further explored by \cite{kubiak_orion_2017}, who combined optical and infrared photometry and identified a rich stellar population towards $\epsilon$ Ori, with an estimated membership of up to $\sim 2300$ stars, known as the Orion Belt Population.

The current view of Orion OB1b emerged from the combination of large spectroscopic follow-up campaigns, Gaia astrometry, and modern clustering techniques. In recent years, precise parallaxes and proper motions from Gaia, together with extensive spectroscopic surveys \citep[e.g.,][]{briceno_cida_2019,kounkel_apogee-2_2018}, have enabled detailed studies of the structure and dynamical evolution of the Orion complex, including OB1b \citep{zari_mapping_2017,zari2019,chen2020,swiggum2021,sanchez2024, sanchez2026}. These studies reveal multiple kinematically distinct populations overlapping along the line of sight towards Orion, making membership assignment and the characterisation of the region particularly challenging.

\cite{kounkel_apogee-2_2018} combined Gaia DR2 photometry and astrometry \citep{gaiadr22018} with radial velocities from SDSS-IV APOGEE-2 \citep[e.g.,][]{ahumada2020} to identify photometric and kinematic candidates towards Orion. They then applied a hierarchical clustering algorithm to a 6D space $(\alpha,\delta,\mu_{\alpha *},\mu_\delta,\varpi,v_r)$ to study the fine structure of the region. They found that the Orion complex is divided into five macro-structures: Orion A, B, C, D, and $\lambda$ Ori. In this framework, OB1b corresponds to a superposition along the line of sight of the Orion C and D populations, including seven substructures: NGC2024, oriCC, $\sigma$ Ori, $\delta$ Ori, $\epsilon$ Ori, $\eta$ Ori, and $\zeta$ Ori. These populations span ages between 1 and 16 Myr and mean distances between 330 and 430 pc.

\cite{briceno_cida_2019}, using a sample of  K- and M-type stars spectroscopically confirmed as young members of OB1b, in combination with Gaia DR2 parallaxes, found a bimodal distance distribution in the region, where a closer population called OB1b I lies to the north of the belt stars, and a farther population called Orion OB1b II lies to the south. Although the authors report a mean age of 5 Myr for the entire OB1b population, they found that the fraction of Classical T Tauri stars (CTTSs) in OB1b I is significantly higher than that of OB1b II, which could suggest that the latter is older.

\cite{zari2019} combined photometric and kinematic information from Gaia DR2 to identify candidate members in Orion, spanning a range of $-2\lesssim M_G \text{ (mag)} \lesssim 10$. Using this sample, they applied the DBSCAN clustering algorithm on the 3D  Galactic Cartesian coordinates to identify sub-structures in the region. They identified seven groups towards OB1b: $E$, $B_2$, $B_3$, $B_4$, $B_5$, $B_7$ and $B_8$ with ages ranging from 4 to 14 Myr and mean distances spanning from $\sim 360$ to $\sim 430$ pc.

\cite{chen2020} applied two unsupervised clustering algorithms to the five-dimensional astrometric space $(\alpha, \delta, \mu_{\alpha^*}, \mu_\delta, \varpi)$ from Gaia DR2, identifying four distinct groups within Orion~OB1b: OBP-near, OBP-b, OBP-d, and OBP-far, with mean heliocentric distances ranging from $360$ to $430$~pc. These groups were independently recovered by \cite{swiggum2021}, who additionally provided age estimates for each subgroup. More recently, \cite{sanchez2024} analysed the Orion complex using a hierarchical clustering approach and distinguished two structural regimes, referred to as large-scale and small-scale structures. Within Orion OB1b, their large-scale groups broadly follow the grouping scheme previously identified by \cite{chen2020} and \cite{swiggum2021}, while their small-scale analysis additionally reveals structures such as OBP-Centre, OBP-South, and OBP-East. The dynamical evolution of these groups was further explored by \cite{sanchez2026}, who accounted for sample incompleteness by assuming a common underlying mass distribution for all groups to initialize their N-body simulations, motivating more comprehensive measurements of the individual IMFs within the region, such as the presented in this work. The projected sky distribution of all groups discussed above is shown in \autoref{figOB1boverview}.

Although the existence of substructure within OB1b is now well established, there is still no consensus on the number of groups, their spatial extent, or their physical nature. The studies discussed above employ different selection criteria for identifying stellar groups, making direct comparisons difficult. This motivated us to perform an independent identification of groups within OB1b in Section \ref{sec:memberships}, which is later used to derive the corresponding mass distributions shown in Section \ref{sec:LandMfunctions}.


\section{Sample of young candidates towards OB1b}\label{sec:candidates}

In this section, we use members and candidates from the literature to characterise the photometric and kinematic properties of young objects towards OB1b and define the selection criteria to construct the Gaia DR3 sample analysed throughout this work.

\subsection{Compendium of known young members and candidates}\label{compendiumsection}

Although the substructure of OB1b remains uncertain, its approximate boundaries on the sky have remained relatively consistent throughout the literature. For our analysis, we considered the region centred on $\epsilon$ Ori ($\alpha=84.05^\circ$, $\delta=-1.202^\circ$) with a radius of $2.5^\circ$, as shown in \autoref{figOB1boverview}, which encloses the areas covered by previous studies.

As a reference catalogue for defining the astrometric and photometric selection criteria, we adopt the NEMESIS catalogue of young stellar objects \citep{roquette2025}, which compiles candidate young stellar objects from a wide range of previous studies. In particular, we focus on the subsample of spectroscopically confirmed members reported by \citet{briceno_cida_2019} (hereafter, the spectroscopic sample), which is included in the compilation. This sample represents a major spectroscopic effort, comprising 2062 spectroscopically confirmed variable sources in the Orion OB1 association, of which 692 are located within our OB1b field. It provides individual measurements of spectral types, $H_\alpha$ emission, and extinction. Although the spectroscopic sample does not cover the full stellar mass range, it is substantially purer than the full NEMESIS catalogue, since its members are identified through spectroscopic signatures of youth. 

\subsection{Flux and mass limits with Gaia DR3}

This work is based on Gaia DR3 \citep{gaia_collaboration_gaia_2022}. Considering only those sources with a full 5D astrometric solution available, we estimated photometric limits by identifying deviations of the logarithmic magnitude distributions from linearity. We converted these magnitudes into masses using the evolutionary models of \cite{baraffe_new_2015}, assuming a mean distance of $400$ pc, an age of $10$ Myr, and a mean extinction of $A_V=0.4$~mag. The resulting limits are $G_{BP}=20.3$~mag ($0.15\,M_\odot$), $G=20.5$~mag ($0.04\,M_\odot$), and $G_{RP}=19.3$~mag ($0.03\,M_\odot$). In particular, for the $G$ and $G_{RP}$ bands the corresponding masses lie within the brown dwarf regime. It is important to note, however, that these photometric limits only ensure detection, but not the precision required to clearly establish kinematic membership. A further refinement of the actual validity ranges of our method is presented in Section~\ref{sec:indmemberships}, where we show that precision of Gaia DR3 astrometry allows for identification of kinematic memberships down to at worst $0.076\,M_\odot$.

\subsection{Kinematic selection}
\label{kinematicselection}

Most recent studies of Orion substructures restrict their analyses to Gaia sources with high-precision parallaxes \citep[e.g.,][]{zari2019,kounkel_apogee-2_2018,sanchez2024}. While this approach strongly reduces field contamination, it also excludes faint candidate members. Since the goal of this work is to determine the IMF over the widest possible mass range, we apply photometric and proper-motion selection criteria to all sources towards OB1b, while distance criteria are applied only to sources with parallax uncertainties below 10\%. Sources without reliable parallaxes are therefore not excluded a priori, but are instead retained based on their photometric and kinematic properties. Since these quantities are generally more precise than parallaxes for faint objects, this approach preserves sources close to the Gaia completeness limits. Distance-dependent analyses involving faint sources, such as the computation of absolute magnitudes, are discussed in Section \ref{sec:LandMfunctions}.

\begin{figure*}[t]
\centering
\includegraphics[width=0.9\linewidth]{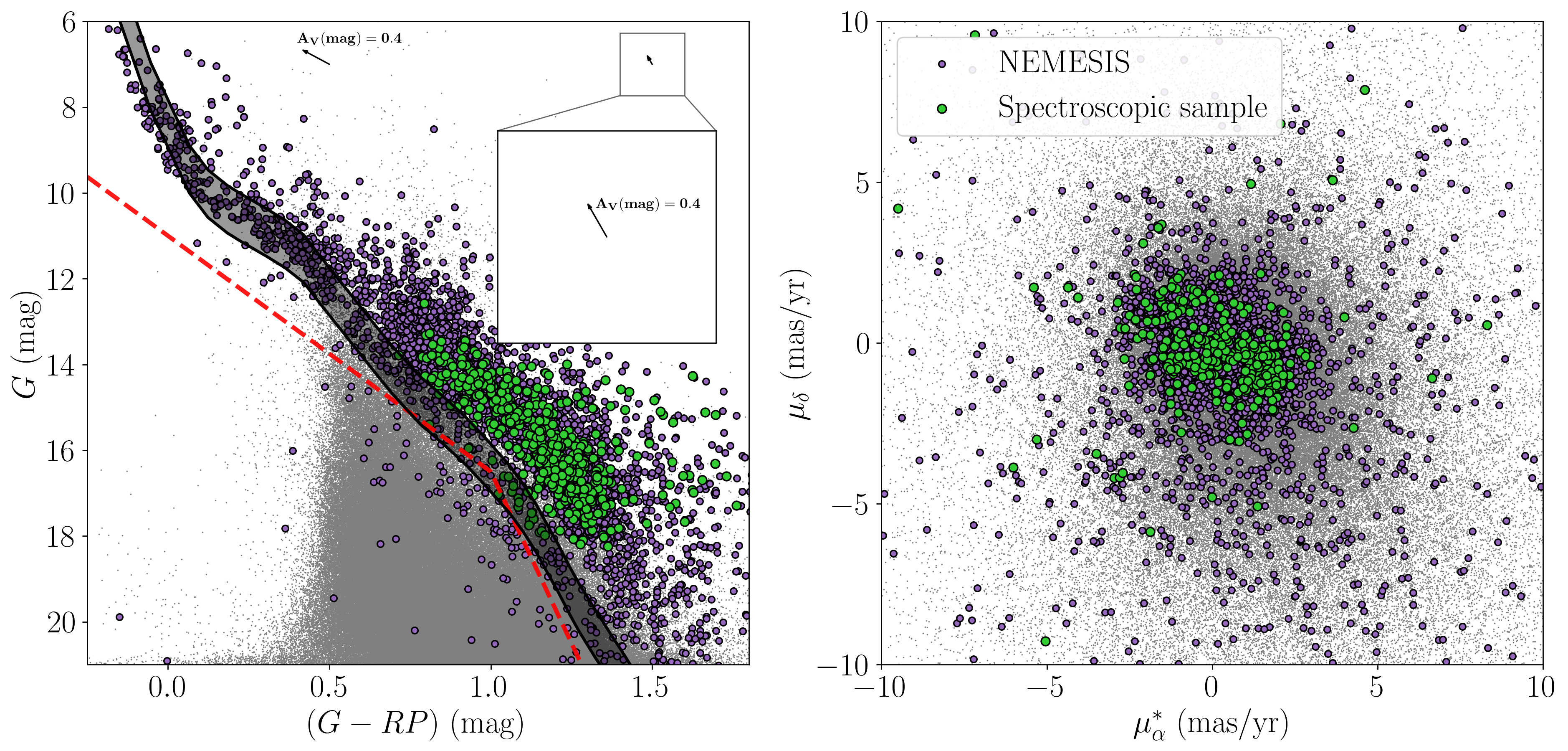}
  \caption{CMD (left) and vector-point diagram (right) 
  of Gaia DR3 sources towards OB1b (grey). Candidate members from NEMESIS 
  are shown in purple and the spectroscopic sample is shown in
  green. The black region in the CMD indicates the area between the solar-metallicity 
  ZAMS at 300 and 500 pc reddened by $A_V=0.4$ mag. Black arrows: reddening 
  vectors at $G-{G_{RP}}=0.5$ and $G-{G_{RP}}=1.5$ mag (shown with a zoom-in box).}
  \label{Fig:Ob1b_photo_kine}
\end{figure*}

The vector-point diagram of the spectroscopic sample (right panel of \autoref{Fig:Ob1b_photo_kine}) shows that most young objects have proper motions satisfying $|\mu_\alpha \cos{\delta}| < 10~\text{mas/yr}$ and $|\mu_\delta| < 10~\text{mas/yr}$. Based on these distributions, we define kinematic candidates as sources within these proper-motion limits. In addition, sources satisfying the parallax measurement precision defined above were also required to have their distances ($d(\mathrm{pc}) = 1000/\varpi(\mathrm{mas})$) to lie between 300 and 500 pc. Applying these selection criteria, we retain 560 out of 692 young sources towards OB1b in the spectroscopic sample. Of the 132 excluded sources, 30 were directly not recovered by Gaia DR3, while the remaining had distances below $300$ pc or above $500$ pc. Although these objects display spectroscopic signatures of youth, they do not appear to be associated with OB1b and could be part of other young populations towards the same direction in the sky.

\subsection{Photometric selection}

Young populations with similar ages and distances define a clear sequence in a colour-magnitude diagram (CMD). Based on the position of objects within the NEMESIS catalogue and the spectroscopic sample, we retained only sources above the red dashed line in the left panel of \autoref{Fig:Ob1b_photo_kine}, that is, all sources satisfying: 

\begin{align}
    \begin{cases}
        G<5.5(G-RP)+11 \quad \mathrm{for}\quad G-RP\leq1~\textrm{mag} \\
        G<16(G-RP)+0.5 \quad \mathrm{for}\quad G-RP>1~\textrm{mag}
    \end{cases}
\end{align}

This selection includes all sources above the solar-metallicity Zero Age Main Sequence at 400 pc reddened by $A_V=0.4$~mag, as well as all members of the spectroscopic sample. Since the selection lines are nearly parallel to the reddening vectors (Section \ref{sec:extinction}), extinction is not expected to move true members outside the selected region. 

Our photometric analysis uses only the $G$ and $G_{RP}$ bands, avoiding $G_{BP}$ because faint low-mass sources are often undetected or have larger photometric uncertainties in this band. The resulting sample after applying the photometric and kinematic selection criteria is hereafter referred to as the science sample and contains 22029 sources.


\section{Memberships to groups in Orion OB1b}\label{sec:memberships}

In this section, we describe the procedure used to identify stellar groups within OB1b and assign individual membership probabilities to candidate sources. First, we introduce the clustering algorithm employed to identify the main groups and compare the resulting structures with those reported in the literature. We then present a probabilistic approach that accounts for astrometric uncertainties to estimate the membership probability of each source to the different stellar groups identified towards OB1b.

\subsection{Population components towards OB1b}
\label{sec:compositepop}

As discussed in Section \ref{sec:OrionOB1b}, recent Gaia-based studies have shown that many young objects towards Orion OB1b are distributed in multiple clustered structures when considering their astrometric properties. However, inspection of the spectroscopic sample in the vector-point diagram (\autoref{Fig:Ob1b_photo_kine}) reveals that a non-negligible fraction of young sources exhibit proper motions inconsistent with the main concentration centred on $\mu\sim0~\mathrm{mas\,yr^{-1}}$, showing instead a broader kinematic distribution. As these objects follow a similar kinematic distribution to the field contaminants, their identification is not straightforward, but we provide a method for their identification in the next sections.

We therefore describe the parameter space towards OB1b as the superposition of three components: $(i)$ a clustered young population identified through clustering algorithms, $(ii)$ a non-clustered young population identified using spectroscopically confirmed members, and $(iii)$ an old field population constrained through the analysis of a control field. The following sections describe the modelling of these components and the probabilistic framework used to estimate individual membership probabilities.

\subsection{Identification of groups with a clustering algorithm}
\label{sec:clustering_algorithm}

We identified the groups within OB1b using the DBSCAN \citep{ester1996} clustering algorithm in the 5D astrometric space $(\alpha,\delta,\mu_\alpha,\mu_\delta,\varpi)$, following the procedure developed by \cite{bertolotto2026}. DBSCAN operates with two free parameters: $\varepsilon$, the maximum separation between neighbouring sources, and $N_{min}$, the minimum number of neighbours required to define a core point. Group identification was performed using only sources with relative parallax uncertainties satisfying $\sigma_\varpi/\varpi<0.1$, reducing contamination from faint objects with uncertain astrometry. Once the groups and their extent are identified, membership probabilities for all sources within the science sample, including those excluded during the clustering stage, are estimated probabilistically in Section \ref{section_faint_members}. Following \cite{bertolotto2026}, DBSCAN was applied in the scaled space $(\alpha/\varepsilon_R,\delta/\varepsilon_R,\mu_{\alpha}/\varepsilon_\mu,\mu_{\delta}/\varepsilon_\mu, \varpi/\varepsilon_\varpi)$, where $\varepsilon_R$, $\varepsilon_\mu$ and $\varepsilon_\varpi$ are the tolerances in sky position, proper motion, and parallax, respectively. 

A fixed value of $\varepsilon=0.763\sqrt{3}$ was adopted, following the procedure described in Appendix E of \cite{bertolotto2026}, while exploring a grid of values for $\varepsilon_R$, $\varepsilon_\mu$, $\varepsilon_\varpi$, and $N_{min}$. The explored parameter ranges were $\varepsilon_R\in[0.05^\circ,0.5^\circ]$ in steps of $0.05^\circ$, $\varepsilon_\mu\in[0.05,0.5]~\mathrm{mas\,yr^{-1}}$ in steps of $0.05~\mathrm{mas\,yr^{-1}}$, $\varepsilon_\varpi\in[0.025,0.25]~\mathrm{mas}$ in steps of $0.025~\mathrm{mas}$, and $N_{min}\in[2,42]$ in steps of 4. To select among the resulting groups, we employed the iterative procedure described by \cite{bertolotto2026}, in which multivariate Gaussian densities are fitted to each group to compute the score function $\mathcal{C}$ defined as:

\begin{equation}
    \mathcal{C}=N\sum _i\frac{f(\mathbf{u_i})}{f(\mathbf{u}_{\text{mean}})}
\end{equation}

where $f(\mathbf{u_i})$ and $f(\mathbf{u}_{\text{mean}})$ are the fitted Gaussian densities evaluated at $\mathbf{u_i}=(\alpha_i,\delta_i,\mu_{\alpha_i},\mu_{\delta_i},\varpi_i)$ and at the mean position of the Gaussian, respectively, and $N$ is the number of objects in the group. The group with the highest $\mathcal{C}$ value is selected, and all overlapping groups are discarded. This procedure 
is repeated until all stars are either assigned to a group or discarded. 

The final sample contains 1869 objects distributed among eight groups, labelled alphabetically from $A$ to $H$, with membership numbers ranging from 91 to 392 (see \autoref{tab:numbersources}). Sources assigned to a cluster via the clustering algorithm are hereafter referred to as clustered candidates, and their distributions among the different groups are shown in \autoref{figclustered_candidates}.

\subsection{Comparison with groups from the literature}
\label{sec:groupcomparison}

Comparing the groups we identified with those reported in the literature is not straightforward, as previous studies employ different clustering methods, selection criteria, astrometric cuts, quality thresholds and membership definitions. Nevertheless, based on their position and compactness on the sky, groups $C$ and $D$ likely correspond to the $\sigma$ Ori and NGC2024 clusters, respectively. NGC2024 is a very young \citep[$\tau\sim1$ Myr, e.g.,][]{robberto2024} and heavily embedded cluster, with extinctions reaching up to $A_V\sim40$ mag. Such extreme extinction makes determining complete IMFs from optical data challenging, and for this reason, group $D$ is excluded from the IMF analysis presented in Section \ref{sec:LandMfunctions}. Similarly, $\sigma$ Ori is also young \citep[$\tau\sim2$ Myr, e.g.,][]{damian2023} and subject to complex differential extinction, making reliable age and mass estimates from optical data alone also challenging. Dedicated analyses of the IMF of $\sigma$ Ori have been carried out by \cite{damian2023} and \cite{pena_ramirez_new_2012}, and since this work is centred on the more dispersed groups of OB1b whose IMFs have not yet been determined, Group $C$ is then excluded as well from our IMF analyses. Nevertheless, both groups are included in the probabilistic membership analysis in Section~\ref{sec:indmemberships}.

\begin{figure*}
    \centering
    \includegraphics[width=0.95\linewidth]{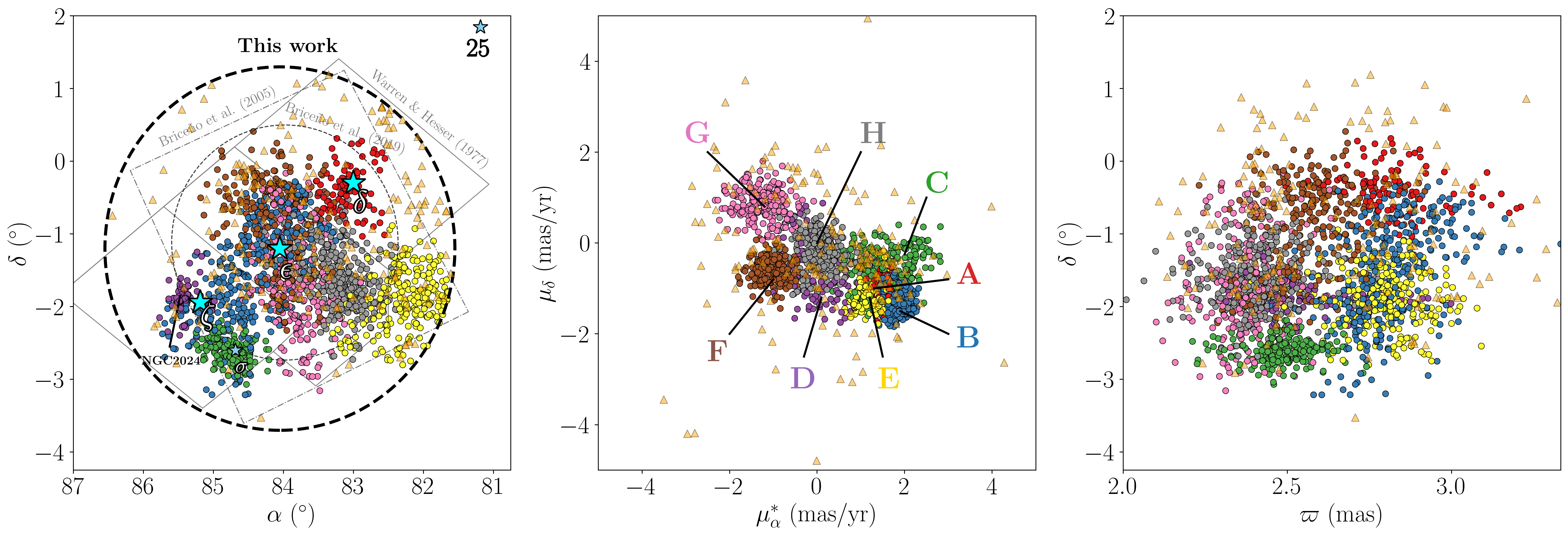}
    \caption{Distribution of clustered candidates within OB1b, colour-coded by group as labelled. Orange triangles: objects from the spectroscopic sample not included by the clustering 
    algorithm. Left: Sky projection. Middle: Vector-point diagram. 
    Right: $\delta$ vs $\varpi$. In the sky projection, the adopted limit for OB1b and 
    limits from the literature are shown.}
    \label{figclustered_candidates}
\end{figure*}

When comparing our groups with OB1b I and OB1b II from \cite{briceno_cida_2019}, defined purely through sky coordinates, we found that they correspond to the superposition of several of our groups. From inspection on the position of our clustered candidates in the sky, we found that objects mostly from groups $A$, $B$, and $F$ compose their 1b I group, while objects mostly from $G$ and $H$ compose 1b II. This correspondence explains the distance bi-modality found by the authors on 1b I, as groups $A$ and $B$ are located much closer than group $F$ (see right panel of \autoref{figclustered_candidates}).

For the analysis of correspondence between groups with other works that also include other astrometric information, we identified correspondences through source-by-source cross-matching between works. Using a sky tolerance of $1''$, we matched our list of clustered candidates with the catalogues from previous studies and analysed how the members of each group are classified in the literature as shown in \autoref{fig:literaturecomparisons}.

In the case of \cite{kounkel_apogee-2_2018}, 779 of the 1184 sources assigned to groups within OB1b are included among our clustered candidates. Because the authors identified a large number of groups, many containing only a few members, our candidates are often associated with multiple of their groups, making direct correspondences less clear. Notable exceptions are groups $E$ and $H$, which largely coincide with the $\eta$-Ori$_1$ and oriCC$_1$ groups, respectively.

A clearer correspondence is found with \cite{chen2020}. Of the 1554 objects toward OB1b in their sample, 711 overlap with our clustered candidates, which are consistently associated with their OBP groups. In particular, members of our groups $A$, $B$, and $E$ are predominantly assigned to the single OBP-near group rather than being separated into distinct structures. \cite{swiggum2021} report 1145 objects toward OB1b, of which 867 are present in our sample. Since they adopt the same grouping scheme as \cite{chen2020}, the correspondence remains largely consistent with the trends described above, except that the OBP-far group is not identified in their work. Similarly, the large-scale structures identified by \cite{sanchez2024} are broadly consistent with the grouping scheme of \cite{chen2020} and \cite{swiggum2021}. Within their small-scale regime, OBP-Centre, a 79-member group, is the only structure entirely contained within our adopted OB1b boundary. Some of its members share sky positions and proper motions similar to those of our groups $F$ and $D$, whereas the remaining members appear more broadly distributed with respect to the groups identified in this work.

Finally, in the case of \cite{zari2019}, 622 of the 938 sources reported towards the Orion Belt are present among our clustered candidates. In general, each of our groups is predominantly associated with a single group in their classification. In particular, groups $A$ and $B$ are mostly identified as part of group $B_3$, while group $E$ is associated with group $B_5$.  Additionally, the distinction between groups $G$ and $H$ is absent, as both are predominantly associated with group E in their work. This is likely because \cite{zari2019} identified groups using only 3D spatial positions, whereas these structures are primarily distinguished by their proper motions (\autoref{figclustered_candidates}).

Overall, we find a broad correspondence between our groups and those reported in the literature. Nevertheless, the distinction between groups $A$, $B$ and $E$ should be interpreted with caution, as these structures are often identified as a single population in some previous studies.

\subsection{Probabilistic approach for individual memberships}
\label{section_faint_members}

Faint sources generally exhibit larger astrometric uncertainties, which can displace them from the loci occupied by cluster members in the 5D astrometric space. For this reason, in Section \ref{sec:clustering_algorithm} we applied DBSCAN only to sources with low parallax uncertainties, obtaining a cleaner identification of the main groups within OB1b. We then used these clustered candidates to model the intrinsic distributions of the groups and estimate membership probabilities for the full science sample.

We assumed that the intrinsic distribution of astrometric parameters of each group is independent of source brightness, equivalent to neglecting mass segregation, so that bright and faint members follow the same underlying distribution. This approximation is supported by previous studies of dispersed Orion populations \citep[e.g.,][]{suarez_system_2019,downes_low-mass_2014}. Under this assumption, we assign membership to all sources in the science sample using a probabilistic mixture model that explicitly accounts for individual astrometric uncertainties. 

The distributions of clustered candidates within each group were modelled using Extreme Deconvolution Gaussian Mixture Models \citep[XDGMMs;][]{bovy2011} in the 5D astrometric space $(\alpha,\delta,\mu_\alpha,\mu_\delta,\varpi)$ using the Python wrapper by \cite{holoien2017}. This method infers the intrinsic distributions deconvolved from astrometric uncertainties by using the full covariance matrix $X_{\rm err}(\mathbf{u})$ of the astrometric parameters $\mathbf{u}=(\alpha,\delta,\mu_\alpha,\mu_\delta,\varpi)$, including their uncertainties and correlations \citep{jaehnig2021}. The full covariance matrix is given in Appendix \ref{app:matrix}. The optimal number of Gaussian components for each group was determined by minimising the Bayesian Information Criterion \citep[BIC;][]{liddle2007}. All groups identified in Section \ref{sec:clustering_algorithm} were adequately described by one or two multivariate Gaussian components.

\subsection{Distribution of the non-clustered population in the 5D space}

To characterise the population of young sources not associated with any of the groups identified by the clustering analysis, we first selected all objects from the spectroscopic sample that were not assigned to any group. We refer to this set as the non-clustered young population of OB1b. Of the 560 young members in the spectroscopic sample, 362 are associated with the clustered populations, while the remaining 198 define the non-clustered component. These objects are shown in \autoref{figclustered_candidates} and exhibit a broader proper-motion distribution than the clustered population. The non-clustered population was modelled as a combination of a uniform distribution over the analysed sky region and four multivariate Gaussian components in the $(\mu_\alpha,\mu_\delta,\varpi)$ space using XDGMMs and BIC.

\subsection{Distribution of the field population in the 5D space}

In addition to young stars, Galactic field sources also populate the parameter space within the Orion OB1b field. To model their distribution, we selected a circular control field with the same angular radius ($2.5^\circ$) as the science sample field, located outside the Orion complex in a region of low extinction. Based on the extinction map of \cite{schlegel1998}, the control field was centred at $\alpha=71.3^\circ$ and $\delta=4.30^\circ$, and all selection criteria described in Section \ref{compendiumsection} were applied to it, yielding a sample of 7112 sources, which reproduce the expected distribution of field stars in the parameter space near OB1b. This distribution is expected to depend on apparent magnitude, unlike that of the groups and the non-clustered population. For instance, faint sources may proportionally have smaller parallaxes than bright sources due to having larger distance moduli. In contrast, the young populations within OB1b would share similar distances regardless of their apparent brightness.

Similarly, the relative proportion of sources of the science sample belonging to the field is also expected to depend on the apparent magnitude. While young sources from OB1b, located at similar distances and having similar ages, would be more frequent in the mass corresponding to the peak of their IMF and less frequent towards the brighter and fainter ends of the brightness distribution. The field, in contrast, results from a combination of multiple populations, spanning a wide range of ages and distances and has an apparent magnitude distribution rising towards the faint end.

To account for this magnitude dependence and enable robust membership probability estimates, we divided the science sample into ten equal subsets sorted by $G$-magnitude. In each bin, we modelled the field distribution using sources from the control field as a combination of a uniform component over the analysed sky region and multivariate Gaussian components in the $(\mu_{\alpha^*}, \mu_\delta, \varpi)$ space. Using BIC, we found that four Gaussian components provided the optimal description of the field distribution in all magnitude bins.

\subsection{The combined distribution}

The distribution of the science sample in the 5D space, $\Phi(\mathbf{u},G)$, is modelled as the sum of three components associated with the clustered groups, the non-clustered young population, and the field:

\begin{align}
 \Phi(\mathbf{u},G)=\Phi_{\text{groups}}(\mathbf{u},G) +\Phi_{\text{non-clustered}}(\mathbf{u},G)+\Phi_{\text{field}}(\mathbf{u},G)
\end{align}

\noindent where:

\begin{align}
    \Phi_{\text{groups}}(\mathbf{u},G)=\sum_{i}^{8}\sum_{j}f_{ij}(G)\mathcal{N}_{ij-\text{groups}}(\mathbf{u}) 
\end{align}

\begin{align}    
    \Phi_{\text{non-clustered}}(\mathbf{u},G)=\frac{1}{A_{sky}}\sum_{j}^{4}g_j(G)\mathcal{N}_{\text{non-clustered}}(\mathbf{u^*})
\end{align}

\begin{align}
    \Phi_{\text{field}}(\mathbf{u},G)=\frac{1}{A_{sky}}\sum_{j}^{4}h_j(G)\mathcal{N}_{\text{field}}(\mathbf{u^*},G)
\end{align}

\begin{figure*}
    \centering
    \includegraphics[width=0.9\linewidth]{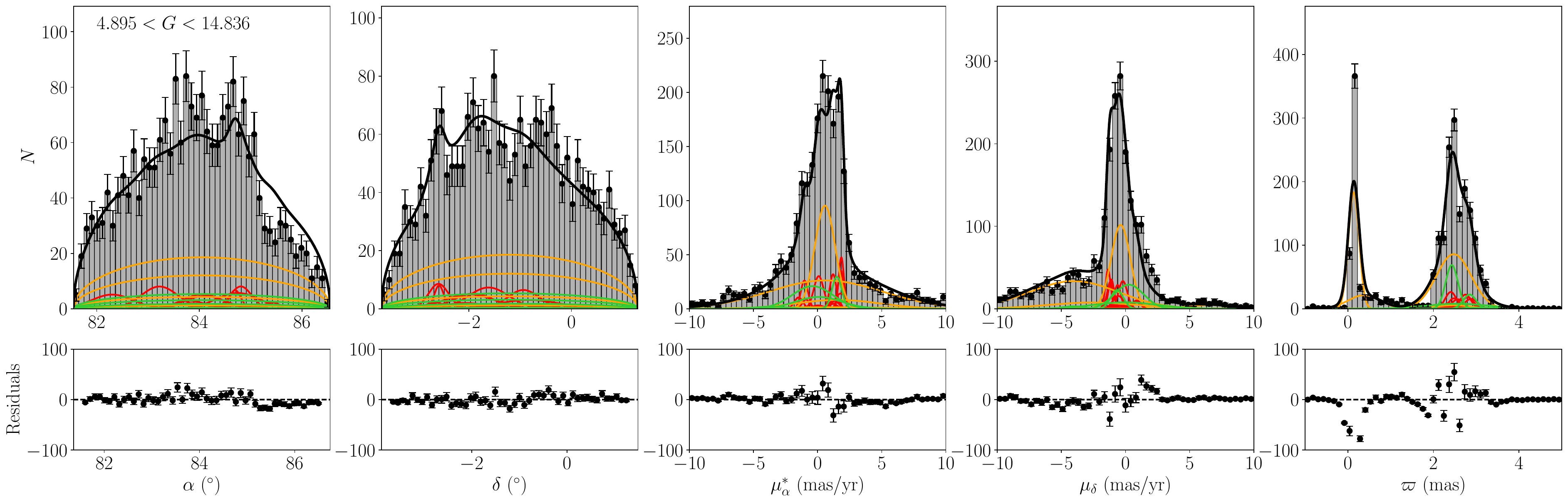}
    \caption{Distributions of the phase-space parameters for sources in the science sample within the magnitude interval 
    $4.895<G~(\mathrm{mag})<14.836$. Black lines: total modelled 
    marginal distributions. Orange, green, and red lines indicate the 
    contributions from the field population, non-clustered population, and clustered groups, respectively. Model distributions were convolved 
    using the mean covariance matrix of sources within the selected $G$-magnitude bin. Bottom panels show the residuals between 
    the observed and modelled distributions.}    
    \label{fig_combined_distributions}
\end{figure*}

where $f_{ij}(G)$ are the weights associated with the $j$-th Gaussian component $\mathcal{N}_{ij\text{-group}}(\mathbf{u})$ of the $i$-th group in the phase space $\mathbf{u}=(\alpha,\delta,\mu_{\alpha},\mu_\delta,\varpi)$; $g_j(G)$ and $h_j(G)$ are the weights associated with the Gaussian components describing the non-clustered population and the field, respectively, in the reduced phase space $\mathbf{u}^*=(\mu_{\alpha},\mu_\delta,\varpi)$; and $A_{\mathrm{sky}}$ is the area of the analysed sky region with radius $2.5^\circ$. We estimated the $f_{ij}$ and $g_k$ coefficients in each magnitude bin using an expectation maximisation algorithm. The marginalised distributions of the phase-space parameters and the corresponding modelled distributions for one representative magnitude bin are shown in \autoref{fig_combined_distributions}. Although the fitting procedure is performed in the full 5D phase space rather than independently on each marginal distribution, the agreement between the observed and modelled marginalised distributions is generally good. In addition, the model distributions shown in \autoref{fig_combined_distributions} were convolved using a single mean covariance matrix for all sources within the selected magnitude bin, whereas the astrometric uncertainties vary from source to source. As a result, small differences between the observed and modelled marginal distributions are expected even for a satisfactory global fit. Nevertheless, the model successfully reproduces the main substructures present in proper motions and parallaxes, supporting the validity of the adopted approach.

\subsection{Individual membership probabilities}
\label{sec:indmemberships}

For each source, we adopted the set of weights corresponding to the 
nearest $G$-magnitude bin. The probabilities of belonging to the $i$-th 
group, the non-clustered population, or the field are then given by 
\citep[e.g.,][]{griggio_astrometric_2022}:

\begin{equation}
    p(\mathbf{u},G)|_{i-\text{th group}}=\frac{\sum_{j}^{}f_{ij}(G)\mathcal{N}_{ij-cluster}(\mathbf{u})}{\Phi(\mathbf{u},G)}
    \label{eqprobabilities1}
\end{equation}

\begin{equation}
    p(\mathbf{u},G)|_{\text{non-clustered}}=\frac{\frac{1}{A_{sky}}\sum_{j}^{4}g_j(G)\mathcal{N}_{\text{non-clustered}}(\mathbf{u^*})}{\Phi(\mathbf{u},G)}
    \label{eqprobabilities2}
\end{equation}

\begin{equation}
    p(\mathbf{u},G)|_{\text{field}}=\frac{\frac{1}{A_{sky}}\sum_j^4 h_j(G)\mathcal{N}_{\text{field}}(\mathbf{u^*},G)}{\Phi(\mathbf{u},G)}
    \label{eqprobabilities3}
\end{equation}

We applied Eqs. \ref{eqprobabilities1}--\ref{eqprobabilities3} to all sources in the science sample and computed membership probabilities for each component. The complete catalogue of sources and their membership probabilities is available in electronic form at the CDS.

Although the implemented mixture model explicitly accounts for individual astrometric uncertainties, very faint sources often have uncertainties large enough to hinder reliable membership assignment. To assess the robustness of our method, we artificially degraded the astrometry of bright sources with high membership probabilities for each group using the covariance matrices of faint sources in the science sample. We found that the membership probabilities remain stable down to $G\sim20$~mag, from which they rapidly decrease as the artificial sources are increasingly assigned to the field component. This behaviour is largely independent of the group considered, leading us to adopt $G=20$~mag as the approximate validity limit of our method. Although fainter sources can still be assigned high membership probabilities to OB1b, their large astrometric uncertainties make these assignments less reliable. The mass associated with this validity limit depends on the age, extinction, and distance of the considered component, and it ranges from 0.032 $M_\odot$ for group $B$ to 0.076 $M_\odot$ for group $G$, with all other values shown in \autoref{tabimfs_orionandbs4}.   

For visualisation and qualitative analyses, we define as most probable candidates of a given group those sources for which the probability of belonging to that group exceeds the probability of belonging to any other component. Nevertheless, all IMF determinations presented in Section \ref{sec:LandMfunctions} make use of the full probabilistic formalism applied to the complete science sample.

The number of sources associated with each component for different probability thresholds is listed in \autoref{tab:numbersources}. The groups contain a number of sources comparable to those identified by the clustering algorithm for probabilities above 50\%, but the number of objects rapidly decreases at higher thresholds, with only a few sources typically exceeding 99\%. This behaviour reflects the strong overlap between groups in phase space (\autoref{figclustered_candidates}). The effect is even more pronounced for the non-clustered population, whose distribution overlaps both with the clustered groups and with the old field. As a result, only 54 sources reach a 99\% membership probability for this component. In contrast, the field population exhibits a slower decline with probability threshold because many field sources occupy regions of phase space with little overlap with the young populations, particularly at large proper motions.

\begin{figure*}
    \centering
    \includegraphics[width=0.9\linewidth]{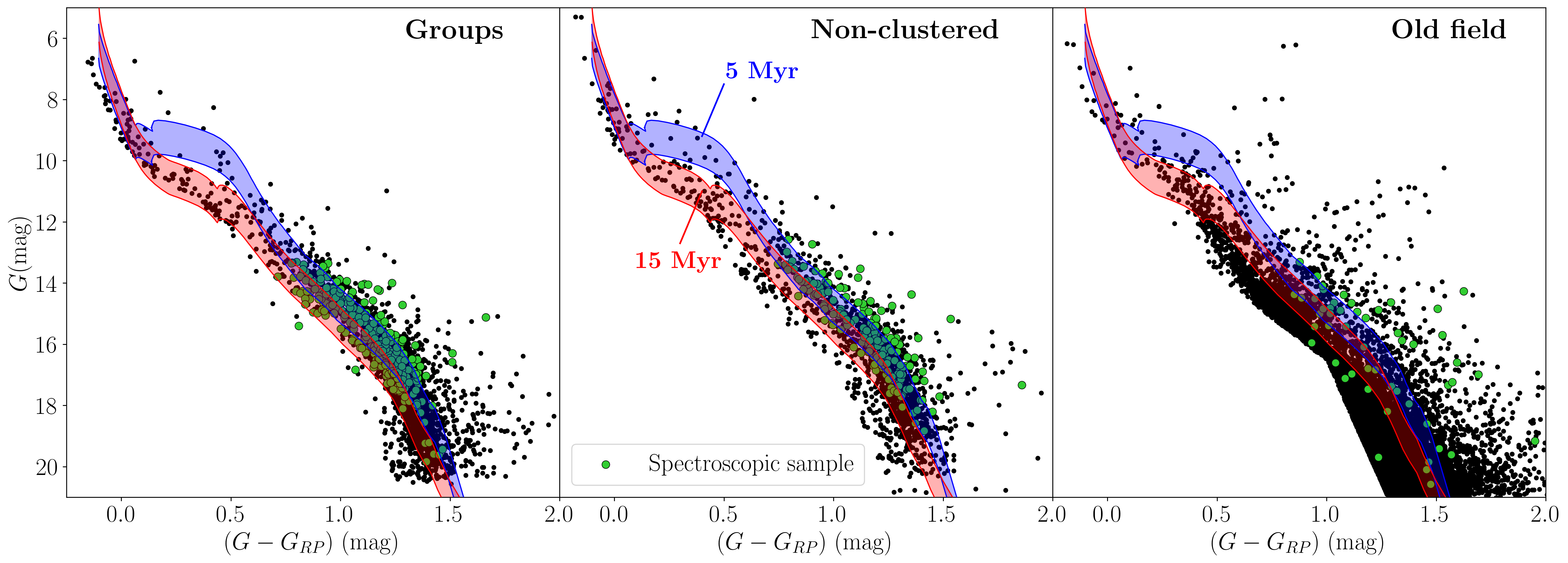}
    \caption{Colour–magnitude diagrams of most probable candidates of the components identified in this work. Left: Clustered groups. Middle: non-clustered population. Right: Old field. Shaded areas: region enclosed by isochrones reddened with $A_V = 0.4$ mag for distances of 300 pc and 500 pc. Blue and red areas correspond to 5 Myr and 15 Myr isochrones, respectively.}
    \label{figCMDdifferent_categories}
\end{figure*}

The distribution in CMDs of the most probable candidates associated with each component is shown in \autoref{figCMDdifferent_categories}. Sources assigned to the clustered populations clearly follow the expected sequence of young stellar populations. In contrast,  sources associated with the old field component do not show a recognisable young sequence, supporting their interpretation as older field contaminants unrelated to Orion. For the non-clustered component, most sources lie within the expected region of a young population; however, some sources are found below the pre-main sequence. This behaviour is expected, as the non-clustered component occupies a more diffuse region of parameter space, similar to that of the old field population, making some level of contamination unavoidable.

The developed probabilistic framework enables deeper comparisons with previous works. In Section~\ref{sec:groupcomparison}, we examined how members of our clustered candidates were classified in the literature. In all cases, cross-matching with our clustered candidates alone did not recover all sources from other works (for instance, only 779 of the 1184 sources from \cite{kounkel_apogee-2_2018} were recovered), reflecting how the choice of clustering algorithm affects the number of sources assigned to a group. Cross-matching instead with our entire science sample (with a 1\arcsec\ tolerance) yields substantially better recoveries. Applying our probabilistic framework to these cross-matched sources, we assigned each one its most probable label among groups, non-clustered, or field.

The resulting label distributions are shown in \autoref{fig:piecharts}. In general, a fraction of the sources assigned to groups in other works are classified as part of the non-clustered component in our framework. The most extreme case is \cite{kounkel_apogee-2_2018}, where 35.7\% of their clustered sources are assigned to the non-clustered component. This likely reflects the finer group partitioning of their work: by identifying a larger number of groups, \cite{kounkel_apogee-2_2018} could subdivide the population more finely, potentially capturing sources that fall between the groups identified in our analysis. Notably, in all comparisons, only a small fraction of sources were classified as field stars by our method, supporting its overall validity.

\begin{table}
    \centering
    \caption{Number of sources with different probability ($p$) thresholds within each component.}
       
    \resizebox{\columnwidth}{!}{\begin{tabular}{ccccc}
    \hline\hline
         Component & Clustered & $p>50\%$ & $p>90\%$ & $p>99\%$  \\
    \hline
         Group $A$ & 101 & 144 & 57 & 0 \\ 
         Group $B$ & 346 & 443 & 312 & 78\\
         Group $C$ & 220 & 242 & 207 & 152 \\
         Group $D$ & 91 & 178 & 106 & 19\\
         Group $E$ & 246 & 272 & 196 & 55 \\
         Group $F$ & 257 & 347 & 241 & 113 \\
         Group $G$ & 216 & 348 & 208 & 30 \\
         Group $H$ & 392 & 457 & 317 & 72 \\
         \hline 
         Non-clustered & --- & 1526 & 645 & 54\\
         \hline
         Old field & --- & 17801 & 16293 & 13465 \\
         \hline\hline
    \end{tabular}}
    \tablefoot{In the case of membership to the groups, the number of objects assigned to each group by the clustering algorithm is also shown.}
    \label{tab:numbersources}
\end{table}

\begin{table*}[]
    \centering
    \caption{Most frequent label assigned to the clustered candidates within each group from this work by other works from the literature, together with their corresponding 
     ages reported by the authors as well as our fits from Section \ref{sec:ages}.}
    \begin{tabular}{cc|cc|cc|cc|cc|cr}
    \hline\hline
\multicolumn{2}{c|}{K+18} &
\multicolumn{2}{c|}{B+19} &
\multicolumn{2}{c|}{Z+19} &
\multicolumn{2}{c|}{C+20} &
\multicolumn{2}{c|}{S+21} &
\multicolumn{2}{c}{This work}\\
\hline
Group   & Age$^a$ & Group & Age$^b$ & Group & Age & Group      & Age & Group & Age & Group & Age \\
\hline
$\delta$ Ori-1       & $4.4\pm1.3$ & 1b I      &  $5.0$   & $B_3$ &  $8^{+0.7}_{-0.04}$    & OBP-near &  \dots   & OBP-near  & $6.8^{+3.6}_{-2.9}$ & $A$ & $5.2^{+0.1}_{-0.1}$ \\
$\zeta$ Ori-2        & $3.3\pm1.6$ &  1b I     & $5.0$   & $B_3$ &   $8^{+0.7}_{-0.04}$   & OBP-near &  \dots   & OBP-near  & $6.8^{+3.6}_{-2.9}$ &  $B$ & $5.2^{+0.1}_{-0.1}$ \\
$\eta~Ori_1$ & $5.5\pm1.4$ &  \dots    &  \dots   & $B_5$ & $8^{+0.7}_{-0.04}$ & OBP-near   &  \dots   & OBP-near  & $6.8^{+3.6}_{-2.9}$ &  $E$ & $6.9^{+0.1}_{-0.1}$ \\
OriCC-2 & $10.0\pm2.4$ &   1b I     &   $5.0$  & $B_7$ & $11.2^{+1}_{-0.1}$ &   OBP-b &   \dots  &  OBP-b  & $17.9^{+6.1}_{-6.5}$ & $F$ & $15.7^{+0.3}_{-0.3}$ \\
OriCC-4 & $8.7\pm1.6$ &  1b II     &  $5.0$   &  $E$ & $11.2^{+1}_{-0.1}$ & OBP-far   & \dots    & ---  & \dots &  $G$ & $15.2^{+0.3}_{-0.3}$ \\
OriCC-1 & $4.6\pm1.4$ &  1b II    &   $5.0$  & $E$ & $11.2^{+1}_{-0.1}$ & OBP-d   &  \dots   & OBP-d  & $6.4^{+6.4}_{-4.3}$ &  $H$ & $8.8^{+0.2}_{-0.2}$ \\
         \hline\hline
    \end{tabular} \\

\tablefoot{
\tablefoottext{a}{The authors report two different ages. The one shown here corresponds to their CMD fitted ages.}
\tablefoottext{b}{The authors report the mean age for the region but argue that 1b I has a higher fraction of Classical T Tauri Stars and suggest that 1b I is younger than 1b II.}}
\tablebib{
(K+18)~\citet{kounkel_apogee-2_2018}; (B+19)~\citet{briceno_cida_2019}; (Z+19)~\citet{zari2019}; (C+20)~\citet{chen2020};  (S+21)~\citet{swiggum2021}
}
\label{tab:GroupAgesLit}
\end{table*}

\section{Luminosity and mass functions}\label{sec:LandMfunctions} 

In this section, we derive the luminosity functions (LFs) and IMFs of the stellar groups and non-clustered population within OB1b using the probabilistic membership formalism described in Section \ref{section_faint_members}. We describe the adopted extinction corrections and the determination of ages required for the assignment of stellar masses. We then present the resulting LFs and IMFs of the different populations identified in the region.

\subsection{Reddening correction and Gaia extinction coefficients}
\label{sec:extinction}

Previous studies of Orion OB1b have adopted different approaches to extinction correction. \cite{kounkel_apogee-2_2018} argued that, for their analysis, extinctions are sufficiently low to be neglected, whereas \cite{zari2019} simultaneously estimated ages and extinctions through isochrone fitting, finding values ranging from $0<A_V<0.5$ mag. In our case, extinction corrections are essential for the determination of stellar masses and the ages of the populations, both of which directly affect the inferred IMFs. 

An important characterisation of extinction across 
OB1b was provided by the spectroscopic sample of \cite{briceno_cida_2019}, who estimated extinctions by comparing observed colours with intrinsic colours inferred from spectroscopically determined spectral types. The sample consists of a large set of 560 stars homogeneously distributed across the region. The extinctions are generally low, ranging from $0<A_V<3$ with a typical value of $\overline{A_V}=0.4$. This sample provides a robust characterisation of the extinction distribution affecting the populations within OB1b. In contrast, although it reproduces well the typical values, the \cite{green2019} 3D dust maps do not reproduce approximately $10\%$ of the sample showing $A_V>1$ mag in the spectroscopic sample, likely because of their limited spatial resolution at the distance of Orion. We therefore de-redenned the photometry of all sources using the spectroscopic extinction distribution. Median extinctions within each group were adopted for age determination, while stochastic extinction values drawn from the distribution were used when constructing the luminosity and mass functions.

Because Gaia passbands are substantially broader than typical optical filters, the extinction coefficient $k_{\mathrm{band}}=A_{\mathrm{band}}/A_V$ depends strongly on stellar colour and extinction. Although \cite{gaia_collaboration_gaia_2022} provides polynomial relations for Gaia EDR3, their calibration spans effective temperatures between 3500 and 10000 K, which does not cover the range expected for the populations studied here. According to the models of \cite{baraffe_new_2015} and PARSEC--COLIBRI \citep{bressan2012,Marigo2017}, our sample should include sources with temperatures between $\sim2600$ K and $\sim25000$ K, corresponding to $\sim0.02~M_\odot$ and $\sim10~M_\odot$ stars at 10 Myr, respectively. Other works have revisited these coefficients \citep[e.g.,][]{ramos2020,monteiro2020,luhmanesplin2020}, but either express the dependencies in terms of intrinsic colours or do not include 
all Gaia passbands.

\begin{figure}[t]
    \centering
    \begin{tabular}{cc}
        \includegraphics[width=0.9\linewidth]{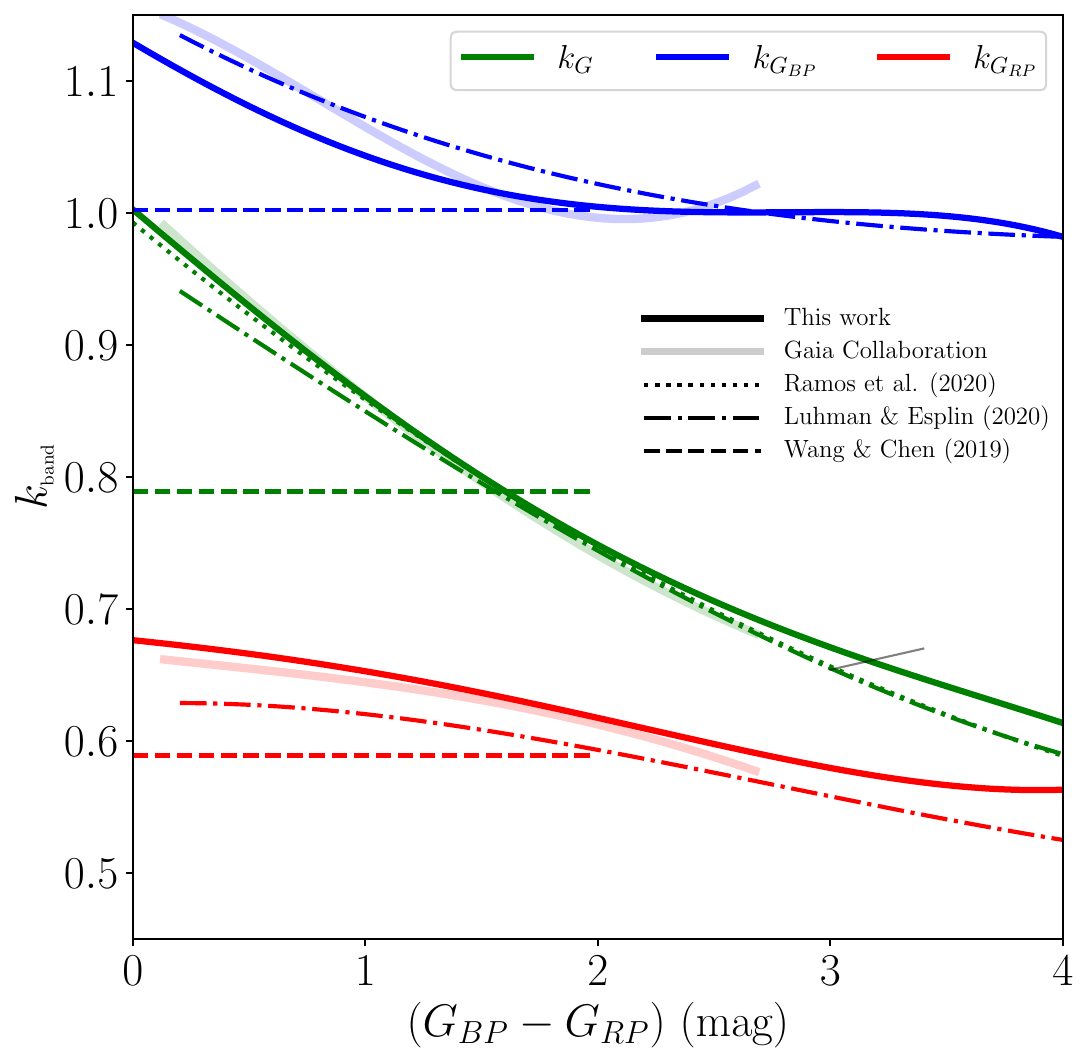} 
    \end{tabular}
    \caption{
    Extinction coefficients $k_{\mathrm{band}}$ as a function of 
    observed $G_{BP}-G_{RP}$ colour and $A_V=0.4$ mag, corresponding to the typical extinction 
    towards OB1b (continuous lines). Values from the 
    literature are shown with different line styles. The colour range shown for \cite{wang_optical_2019} corresponds to the approximate validity range adopted in their work.}
    \label{figextinctions}
\end{figure}

Following \cite{Danielski2018}, we recalculated the colour-dependent extinction coefficients through the integration of reddened spectra from the BT-NextGen models \citep{allard2011} with $T_{\rm eff}$ between 2600 and 30000 K and $\log g=4.25$, typical of pre-main-sequence stars. The spectra were reddened using the \cite{fitzpatrick2019} extinction law with $A_V$ values between 0.1 and 5 mag in steps of 0.1 mag, and convolved with the Gaia passbands from \cite{maiz2019}. The extinction
coefficients were then computed as:

\begin{equation}
 k_{band}=\frac{A_{band}}{A_V}=
 -\frac{2.5}{A_V}\log_{10}\left(
 \frac{\int F_\lambda T_{\lambda,band}E_{\lambda}^{A_V}d\lambda}
 {\int F_\lambda T_{\lambda,band}d\lambda}
 \right)
\end{equation}

where $E_{\lambda}^{A_V}$ is the extinction function from \cite{fitzpatrick2019}, $T_{\lambda,\mathrm{band}}$ is the transmission curve of the corresponding Gaia passband, and $F_\lambda$ is the spectral flux. A fourth-order polynomial in colour and $A_V$ was then fitted to $k_{\mathrm{band}}$, i.e.:

\begin{equation}\label{eqexcfit}
k_{band}(x, A_V)=
\sum_{i=0}^{4}\sum_{j=0}^{4-i} c_{ij}~x^i A_V^j 
\end{equation}

where $x$ is either $G-RP$ or $BP-RP$ and $c_{ij}$ are the fitting coefficients. 
All coefficients obtained from the fit are listed in \autoref{tab_extinction_coefficients}, and \autoref{figextinctions} shows a comparison with literature relations. 
Our relations agree well with the results of \cite{ramos2020}, \cite{luhmanesplin2020}, and the Gaia Collaboration for the $k_G$ coefficient, while some slight differences arise in the colour dependence of $k_{G_{BP}}$ and $k_{G_{RP}}$. Our $k_{G_{RP}}$ estimates follow the same trend as those of \cite{luhmanesplin2020}, but yield larger values across the entire colour range. For $k_{G_{BP}}$, our estimates are smaller at the blue end and similar to those of \cite{luhmanesplin2020} at the red end. The Gaia Collaboration reports a similar trend, with slight differences at the blue end, but their coefficient begins to increase towards redder colours, a behaviour not seen in either our estimates or those of \cite{luhmanesplin2020}. Overall, despite the mentioned differences, the general trends of the coefficients' dependence on colour are in good agreement between works, despite being obtained from different extinction laws and stellar spectra models. Finally, the single-value coefficients reported by \cite{wang_optical_2019} provide good approximations to our $k_{G_{BP}}$ and $k_{G_{RP}}$ estimates, but yield only an intermediate value for $k_G$, as the strong dependence on colour for the coefficient is not included in their work.

\subsection{Multiplicity}
\label{sec:multiplicity}

This work is centred on the determination of the system IMF, that is, the IMF of unresolved stellar systems \citep{chabrier_galactic_2003} treated as single sources when deriving masses from photometry. This approach has been widely adopted in studies of the IMF in Orion \citep[e.g.,][]{bayo_spectroscopy_2011,pena_ramirez_new_2012,da_rio_initial_2012, downes_low-mass_2014,suarez_system_2019}. Under the assumption that binarity properties are similar across different populations, this approach allows meaningful comparisons between system IMFs. A robust determination of the underlying single-star IMF in OB1b is not currently achievable with available observations and is therefore not addressed in this work.

\subsection{Ages by isochrone fitting}
\label{sec:ages}

As discussed in Section \ref{sec:OrionOB1b}, the stellar groups within 
OB1b span a wide range of ages. Since the conversion from photometry to stellar masses depends strongly on age, assuming a single representative age for all groups could bias the derived IMFs. We therefore estimated ages independently for each group through isochrone fitting following an analogous method to the one employed by \cite{zari2019}. For each source, the likelihood of it coming from an isochrone with age $\tau$ and extinction $A_V$ is given by:

\begin{align}
&L(\mathbf{q}^{obs}~|~\tau, m)=\prod_{i=1}^2\left(\frac{1}{(2 \pi)^{1 / 2} \sigma_i}\right) \times e^{-\frac{(q_i^{obs}-q_i(\tau,A_{V},m))^2}{2\sigma_i^2}}
\label{likelihood}
\end{align}

where $\mathbf{q}^{\text{obs}}$ and $\mathbf{q}(\tau,A_V,\varpi,m)$ are the vectors of observed and modelled quantities respectively. In this case $\mathbf{q}(\tau,A_V,\varpi,m)=(G_{mod},(G-RP)_{mod})$, with:

\begin{align}
    G_{mod}= & M_G(\tau,m)-5\notag  + 5\log(1000/\overline{\varpi}) \notag \\ + & k_{G|(G-RP)_{obs}}\times \overline{A_V}
\end{align}

\begin{align}
     (G-RP)_{mod}= & M_G(\tau,m)-M_{RP}(\tau,m)\notag +k_{G|(G-RP)_{obs}}\times \overline{A_V} \notag \\ - & k_{RP|(G-RP)_{obs}}\times \overline{A_V}
\end{align}

where $\overline{\varpi}$ and $\overline{A_V}$ are the median parallax (in mas) and extinction of clustered candidates within the considered group respectively, $(G-RP)_{obs}$ is the observed colour of the source whose likelihood is being calculated, and $k_{G|(G-RP)_{obs}}$ and $k_{RP|(G-RP)_{obs}}$ are the $G$ and $G_{RP}$ extinction coefficients associated to the observed colour according to \autoref{eqexcfit}. The computation of the absolute magnitudes $M_G$ and $M_{RP}$ is done by linearly interpolating within PARSEC-COLIBRI (for $m>1M_\odot$) and \cite{baraffe_new_2015} (for $m<1M_\odot$) solar metallicity isochrones.
    
To account for the non-uniform density of sources along the isochrones, 
following \cite{zari2019}, we weighted the likelihood of the $j$-th 
source by $w_j$, where:

\begin{align}
    w_j=\sqrt{n_{redder ~j}/(n_{bluer~j}+1)}
\end{align}

where $n_{redder}$ is the number of stars with $G-RP$ greater than that of  the $j$th star and $n_{bluer}$ the number of stars with $G-RP$ smaller than that of the $j$th star. The likelihood for $N$ coeval stars is then defined as:
\begin{align}
    L_{\text {combined }}(\tau, m)=\prod_{j=1}^N L_j(\mathbf{q}^{obs}~|~\tau, m)^{w_j}
\end{align}
Finally, to get the likelihood depending only on age, we integrate on $m$, resulting in

\begin{align}
    L_{\text {combined}}\left(\tau\right)= \int L_{\text{combined}}\left(\tau, m\right) \mathrm{d} m
\end{align}

The adopted age of each group corresponds to the value maximising $L_{\mathrm{combined}}(\tau)$, while the uncertainties were estimated from the 16th and 84th percentiles of the likelihood distribution. The best-fitting isochrones for all groups are shown in the CMDs of \autoref{figCMDallgroups}. The resulting ages together with values from the literature are summarised in \autoref{tab:GroupAgesLit} and range from $\sim5.2$ Myr (groups $A$ and $B$) to $\sim15.7$ Myr (group $F$). Since groups $A$ and $B$ exhibit essentially the same fitted ages, they are consistently associated with the OBP-near population in previous works \citep{chen2020, swiggum2021}, they are located adjacent to each other in the sky, and have very similar proper motions and parallaxes, we argue that they may be part of the same bigger structure (from now on $A+B$). In contrast, group $E$ seems to be older ($\tau\sim6.9$ Myr), suggesting that it may represent a distinct population despite also being associated with OBP-near in the literature. Although some overlap in age can also be found between groups $F$ and $G$, their clear difference in proper motions makes it difficult to argue that they belong to the same structure. Nevertheless, establishing the physical relation between these groups is beyond the scope of this work, and all populations are therefore treated independently for the IMFs determinations.

Our age estimates generally align with the typical ages assigned to the different OB1b groups in the literature. In the case of \cite{briceno_cida_2019}, although they report an average age for the region of 5 Myr, the authors argue that their 1b\,I and 1b\,II groups may differ in age, based on their different fractions of Classical T Tauri Stars (CTTSs), with 1b I having a considerably larger fraction. This is consistent with our findings: 1b I is predominantly composed of our groups $A$, $B$, and $F$, spanning a wide age range from $\tau \sim 5$ Myr (groups $A$ and $B$) to $\tau \sim 16$ Myr (group $F$), while 1b II is mostly composed of groups $G$ ($\tau \sim 15$ Myr) and $H$ ($\tau \sim 9$ Myr). Despite both subregions containing older components, the presence of the young groups $A$ and $B$ in 1b I makes it, on average, younger than 1b II, supporting the CTTS fraction argument of \cite{briceno_cida_2019}.

\subsection{Consistency of age trend with near-infrared excesses}

As a consistency check of the age trend inferred from isochrone fitting, we estimated the fraction of sources in each group showing near-infrared excesses which are commonly used as tracers of circumstellar discs, and whose frequency decreases with age \citep[e.g.,][]{hernandez2007}. For this purpose, we constructed $K-W_2$ versus $z-J$ colour-colour diagrams for each group (\autoref{figCCallclusters}) with photometry from 2MASS \citep{skrutskie2006}, WISE \citep{wright2010} and Pan-STARRS \citep{chambers2016}. 

Following an analogous infrared excess criterion as that adopted by \cite[][see Figure 1]{luhman2012}, we classified as infrared-excess sources lying more than 0.2 mag above the expected $K-W_2$ colour from the isochrone corresponding to the age of each group. Then, the infrared-excess number fraction $f_{exc}$ was computed as the probability-weighted number of sources showing infrared excesses divided by the probability-weighted total number of objects in the group. For each group, we compute $f_{exc}$ for the science sample and the spectroscopic sample. Since the fraction of objects with infrared excesses is expected to change with mass \citep[e.g.][]{downes_number_2015}, it is also sensitive to the considered colour range, as colour also correlates with mass. For a fair comparison with the literature, we estimated these fractions only considering the colour range $1<z-J~\mathrm{(mag)}<1.9$, which includes the majority of objects in the spectroscopic sample and encompasses spectral types ranging from $\sim$ K0 to $\sim$ M6.

\begin{figure}
    \centering
    \includegraphics[width=0.9\linewidth]{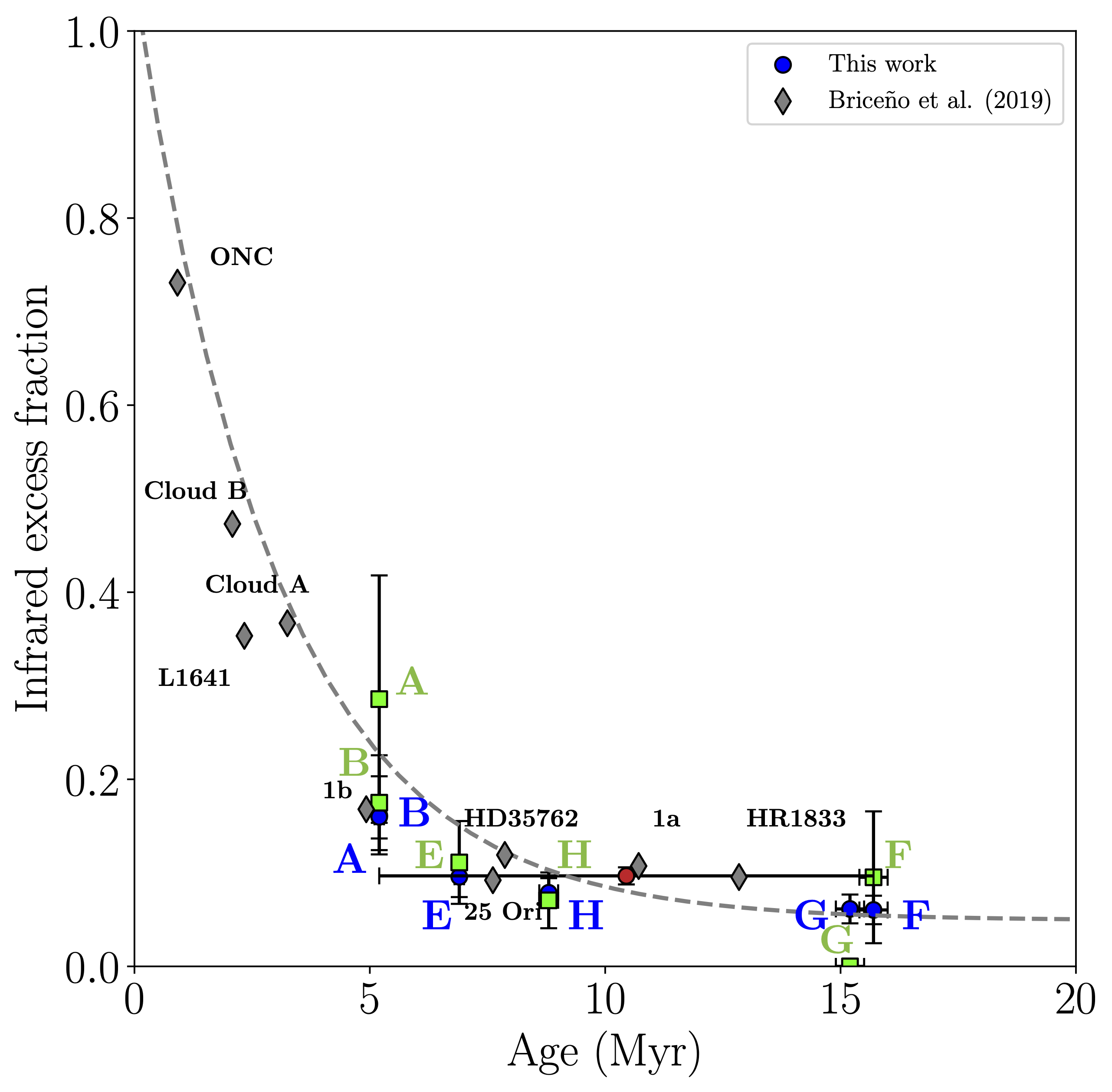}
    \caption{Fraction of sources showing infrared excesses in the $K-W2$ 
    colour as a function of the age within each group (indicated with blue labels) and the non-clustered component as a function of age. Circles: considering all sources in the science sample with photometric colours satisfying $1<z-J<1.9$ in the groups (blue) and the non-clustered component (red). Green squares: considering only sources within the spectroscopic sample. Grey rhombuses: Infrared excess fraction for different regions in Orion according to \cite{briceno_cida_2019}. The dashed grey line indicates the exponential fit found by \cite{briceno_cida_2019} in their sample of populations. }
    \label{figinfrared_excesses}
\end{figure}

As discs causing these excesses would dissipate over time \citep[e.g.,][]{briceno_cida_2019}, an exponential decay in the fraction of sources with infrared excesses is expected. In \autoref{figinfrared_excesses} we show the dependency of this fraction on age for the different groups in this work, as well as the results for other regions of Orion as reported by \cite{briceno_cida_2019}. When comparing our findings with the exponential fit found by the authors, we found clear consistency with other populations. The similar infrared-excess fractions obtained for the science and spectroscopic samples indicate that our kinematically selected sample traces a population consistent with that identified through spectroscopy. As infrared excess is independent of the membership determination, this provides an independent validation of our probabilistic membership method and supports the identification of additional members not included in previous spectroscopic studies of the region.

\subsection{The probability-weighted luminosity functions}
\label{sectionLFs}

We computed the absolute $G$-band LF of each group by estimating the absolute magnitudes $M_G$ of all sources in the science sample and weighting their contribution to each group by the corresponding membership probabilities. Absolute magnitudes were computed as:

\begin{equation}
M_G = G + 5 - 5 \log \left( 1000/\varpi (\text{mas})\right) - k_{G|G-RP}\times A_V
\end{equation}

Since spectroscopic measurements of $A_V$ are not available for all sources, extinction values were stochastically drawn from the spectroscopic $A_V$ distribution associated with each group and converted into $A_G$ using the extinction coefficients derived in Section \ref{sec:extinction}.

To account for parallax uncertainties, we drew parallaxes for each source from Gaussian distributions centred on the measured Gaia parallaxes with standard deviations equal to the reported uncertainties. This procedure was repeated 1000 times. In each realisation, the LF was constructed by weighting each source by its membership probability to the corresponding group and binning the resulting $M_G$ values between $-2$ and $14$ mag using 1-mag bins. The reported luminosity function for each group corresponds to the mean of all realisations. Uncertainties were computed as the sum in quadrature of the standard deviation among realisations and the Poisson uncertainty. 

To evaluate the impact of low-probability objects on the resulting LFs, we repeated the process using only the most probable candidates as defined in \autoref{sec:indmemberships}. \autoref{figlfsgroups} shows the resulting LFs, which are broadly consistent with those obtained from the complete science sample. As expected, the main differences appear at the faint end, where larger uncertainties lead to membership probabilities being distributed among multiple groups. Although the individual assignments of these sources are more uncertain, the probabilistic weight scheme makes the LFs derived from the complete science sample more representative of each group.

\subsection{The probability-weighted IMF of the groups}
\label{sectionIMFs}

\begin{figure*}
    \centering
    \includegraphics[width=0.95\linewidth]{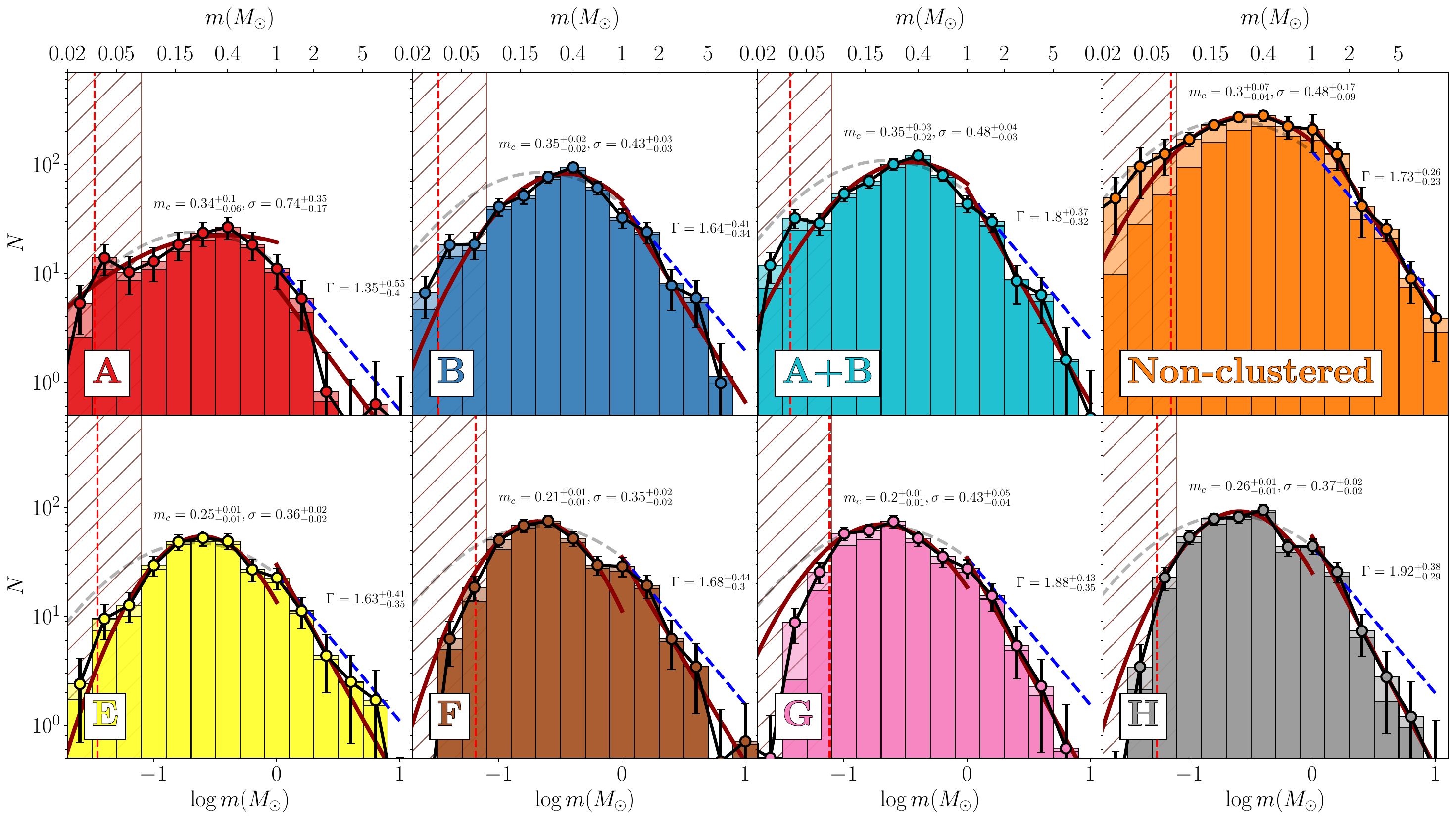}
    \caption{Probability-weighted IMFs of all groups and the non-clustered population within 
    Orion OB1b, derived using the complete science sample (light
    colours) and the most probable candidates (dark colours). 
    The \cite{salpeter_luminosity_1955} and \cite{chabrier_galactic_2003} 
    parameterisations are shown in grey and blue, respectively. Red continuous 
    lines show the best-fitting lognormal models for $m<1~M_\odot$ and the best-fitting power-law for $m>1~M_\odot$. The best-fitting lognormal and power-law parameters $m_c,\sigma,\Gamma$ are also shown. The shaded brown region marks the brown dwarf regime ($m<0.08~M_\odot$),
    and the vertical dashed line indicates the mass corresponding to 
    the approximate validity limit of G=20~mag (Section \ref{sec:indmemberships}).}
    \label{figimfsgroups}
\end{figure*}

Using the same probabilistic framework adopted for the luminosity functions, we computed the IMF of each group by converting the photometry of all sources in the science sample into stellar masses and weighting their contribution by the corresponding membership probabilities. For each group, we assumed the age and corresponding uncertainty derived in \autoref{sec:ages}, with parallaxes and extinctions treated as described in \autoref{sectionLFs}. We generated 1000 realisations of the mass calculation by estimating stellar masses through maximisation of the likelihood defined in \autoref{likelihood}. Realisations differ from one another through the differences in the assigned ages, parallaxes, and extinctions and in each one of them the IMF was constructed by weighting the contribution of each source by its membership probability and considering logarithmic bins ranging from $0.02~M_\odot$ to $12.7~M_\odot$ with widths of 0.2 dex, so that the sub-stellar regime ($m<0.08~M_\odot$) is included in the first three bins. Finally, the reported IMF of the group is the mean of all realisations, and the uncertainty as sum in quadrature of the standard deviation across all realisations and the Poisson uncertainty. The results are shown in \autoref{figimfsgroups} together with canonical parametrisations and the corresponding best-fitting models. The same analysis, repeated for the most probable candidates, revealed trends fully consistent with those found for the LFs.

The resulting IMFs were fitted using three standard parameterisations only considering the bins above the mass corresponding to the validity limit of our method for each group: a lognormal distribution for low masses (${m<1~M_\odot}$), a single power law for high masses (${m>1~M_\odot}$), and a tapered power law over the full valid mass range. The adopted functional forms are described in Appendix \ref{sec:parametrisations}, and the fitted parameters for all groups are listed in \autoref{tabimfs_orionandbs4}. The fitting procedure was repeated 1000 times, with different stochastic realisations of the IMF generated by re-sampling each bin value from a Gaussian distribution centred on the reported value, with a standard deviation equal to the reported uncertainty. The reported best-fitting parameters were then taken as the median across all stochastic realisations, and the uncertainties were reported as the 84th and 16th percentiles. For $m<1~M_\odot$, the lognormal fits generally yield larger characteristic masses $m_c$ and smaller dispersions $\sigma$ than the canonical Galactic-field values of $m_c=0.2~M_\odot$ and $\sigma=0.6$ reported by \cite{chabrier_galactic_2003}. For $m>1~M_\odot$, most groups are indistinguishable within uncertainties from the \cite{salpeter_luminosity_1955} slope of $\Gamma=1.35$, although the reported median value is consistently steeper.

\subsection{The probability-weighted IMF of the non-clustered population}
\label{imfnonclustered}
Because the non-clustered population is not associated with any single group, assigning representative ages for IMF determination is not straightforward. Assuming that this component is composed of both outliers from the clustered populations and stars dynamically dispersed from them, we adopted as a rough approximation the same age distribution as that of the clustered groups. This distribution was constructed from the most probable candidates of all groups, excluding groups $C$ and $D$, assigning to each source the age of its corresponding group.

The IMF was then computed following the same procedure used for the groups. In each realisation, parallaxes were drawn from Gaussian distributions centred on the measured Gaia parallaxes, extinction values were drawn from the spectroscopic $A_V$ distribution, and ages were drawn from the age distribution described above. Stellar masses were estimated through maximisation of the likelihood defined in \autoref{likelihood}, and the resulting mass histogram was weighted by the membership probability to the non-clustered component. This procedure was repeated 1000 times, and the reported IMF corresponds to the mean of all realisations. For comparison, the same calculation was also performed 
using only the most probable candidates.

The resulting IMF is shown in \autoref{figimfsgroups}. Its overall shape peaks at higher masses than the Galactic-field IMF. At higher masses, it follows a power law consistent within uncertainties with the \cite{salpeter_luminosity_1955} canonical IMF. At the low-mass end, the non-clustered population IMF is shallower than those of the groups. However, the significantly smaller number of objects in the IMF derived from the most probable candidates suggests that contamination may partially contribute to this flattening. If the difference is instead physical, it could reflect differences in the dynamical evolution of low-mass objects. Addressing the physical origin of this discrepancy, however, would require dedicated N-body simulations and lies beyond the 
scope of this work.

\subsection{Gravitational state of the groups}

Most young stellar groups disperse within their first $\sim10$ Myr, while only a small fraction survives for $\sim100$ Myr or longer \citep[e.g.,][]{lada_2003}. To investigate the dynamical state of the groups identified in OB1b, we computed the lower bound of the virial ratio, $Q_{\mathrm{min}}=2T_{\mathrm{min}}/|U|$, where $T_{\mathrm{min}}$ is the kinetic energy estimated using only tangential velocities, and $U$ is the gravitational potential energy. Using the derived membership probabilities, these quantities can be estimated for the $k$-th group as:

\begin{align}
    &T_{min}^{k}=\frac{1}{2}\sum_i p_i^km_i(v_{\alpha i}^2+v_{\delta i}^2) \label{eqkineticenergy}\\
    &U^{k}=-G\sum_{i<j}\frac{p_i^kp_j^km_im_j}{r_{ij}} \label{eqpotentialenergy}
\end{align}

where the sum extends over all objects in the science sample and $p_i^k$ is the membership probability of the $i$-th star of belonging to the $k$-th group. The masses $m_i$ were computed as explained in Section \ref{sectionIMFs}. The resulting values range $10\lesssim Q_{min}\lesssim 9.2\times10^{2}$ for all groups. Since $T_{min}$ provides only a lower bound to the true kinetic energy, $Q_{min}$ provides a lower bound of the actual virial ratio $Q$. The inferred values $Q\gg1$ therefore support that all groups are gravitationally unbound.

Because the computation of $Q_{\mathrm{min}}$ does not account for astrometric uncertainties, which could artificially increase the velocity dispersion, we performed a simple numerical test to estimate their impact. We generated synthetic virialised groups with members at the same positions as the real ones, but with velocities chosen so that $Q = 1$. To do this, we drew 3D velocities from a normal distribution centred at zero with $\sigma = 1$ km/s, computed the total kinetic and potential energies $T$ and $U$, and scaled all velocities by $\sqrt{|U|/2T}$ to force $Q = 1$. We then replaced two components of the 3D velocity with realisations that account for the expected proper motion uncertainties of sources at different magnitudes within each group. Repeating this process 1000 times and computing $Q_{min}$ for each realisation, we find that the resulting values never exceed 30\% of the measured ones, which confirms the conclusion that all groups are indeed gravitationally unbound.

\section{Discussion}
\label{sec:discussion}

The main goal of this work is the determination of the IMF of Orion OB1b and its different stellar populations over the mass range accessible with Gaia DR3. In this section, we discuss how the IMFs derived here compare with each other and with different populations identified within OB1b and with previous IMF determinations across the Orion star-forming region.

\subsection{Consistency between the IMFs within OB1b}
\label{sec:OrionOB1bIMF}

At first sight, the IMFs derived for the OB1b groups suggest a broad division into two sets: groups $A$ and $B$, and groups $E$, $F$, $G$, and $H$. Groups $E$, $F$, $G$, and $H$ exhibit IMFs consistent with log-normal distributions for $m<1\,M_\odot$, with characteristic masses of $m_c\sim0.2\,M_\odot$ and widths $\sigma\sim0.4$. In contrast, groups $A$ and $B$ peak at higher masses ($m_c\sim0.4\,M_\odot$) and show broader low-mass distributions ($\sigma\sim0.6$), a behaviour also observed in the non-clustered population. The main differences are therefore confined to the low-mass regime, while the high-mass slopes $\Gamma$ tend to be slightly steeper than the canonical Salpeter value ($\Gamma=1.35$), although they remain statistically consistent within uncertainties.

To quantify these differences, we performed pairwise Kolmogorov-Smirnov (KS) tests across all groups using their measured IMFs over the mass range in which they are valid ($m>0.076~M_\odot$), with the resulting $p$-values shown in \autoref{figkstest}. We adopt a significance threshold of $p<0.05$ to reject the null hypothesis that two samples are drawn from the same underlying distribution.

In the high-mass regime ($m>1M_\odot$), the results show consistency among all groups with $0.309<p<0.997$ although the small sample sizes in this mass range require these results to be interpreted with caution. When considering the full valid range ($m>0.076M_\odot$), the results show no rejected pairs among groups $E$, $F$, $G$ and $H$. The comparison of group $A$ with all remaining groups yields values in the range $0.101<p<0.927$, and therefore does not reject consistency with any group. This result should nevertheless be interpreted with caution, as the small sample size of group $A$ reduces the statistical power of the test. Group $B$ shows $p=0.061$ and $p=0.086$ when compared to groups $E$ and $H$, respectively, and $p\ll0.001$ when compared to both groups $F$ and $G$. These results indicate the IMF of $B$ is statistically inconsistent with groups $E$ and $H$.

Nevertheless, groups $A$ and $B$ deserve particular attention. As discussed above, they exhibit very similar proper motions and parallaxes while occupying adjacent regions on the sky. This, together with their coeval ages, suggests that they may form a single extended structure spanning much of OB1b. If so, their large spatial extent and overlap with other components may complicate membership assignment, allowing a non-negligible fraction of their members to be classified as part of the non-clustered component. Motivated by their possible interpretation as a single coeval structure, we also tested the combined IMF of $A+B$, which is consistent with groups $E$ ($p=0.066$) and $H$ ($p=0.088$) while remaining inconsistent with groups $F$ and $G$ ($p\ll0.001$).

Finally, when comparing the IMF of the non-clustered component with those of the individual groups, we find statistical inconsistency with all groups ($p<0.017$) except $A$ ($p=0.117$). Since the determination of the non-clustered IMF is subject to the limitations outlined in Section \ref{imfnonclustered}, interpreting these comparisons should be done with caution. 

Overall, these results suggest that most OB1b groups are consistent with a common IMF, while the apparent deviations observed in groups $A$, $B$, and the non-clustered component are likely influenced by their more complex spatial and kinematic structure. Although genuine IMF differences cannot be excluded, the present evidence indicates that methodological effects related to membership assignment may play an important role in shaping these discrepancies. Future radial-velocity measurements, particularly for the least massive members of these components, will be essential to improve membership determinations and clarify whether the observed differences reflect intrinsic IMF variations.

\subsection{The combined IMF of OB1b}

We explore the net stellar mass distribution of OB1b as a whole by adding together the individual IMFs of the clustered and non-clustered populations. This combination is justified by the age obtained for the groups, which prevents differential evolutionary effects at the high-mass end from dominating the summed IMF. The resulting combined IMF is shown in the left panel of \autoref{fig:alphaplotandIMF}, together with the summed IMFs of the clustered groups, the IMF from the non-clustered population, and the canonical Galactic-field parametrisations. Integrating the probability-weighted source counts over the resulting mass bins yields 2044 sources in the clustered component and 1892 in the non-clustered component, corresponding to a total of 3936 young sources in OB1b.  

For the combined IMF, we recover a distribution well described by a log-normal function for $m<1~M_\odot$, with $m_c= 0.27^{+0.02}_{-0.02}$ and $\sigma=0.45^{+0.03}_{-0.03}$, and by a power-law for $m>1~M_\odot$, with $\Gamma = 1.85^{+0.16}_{-0.14}$. These values are consistent with those derived for the combined groups ($m_c=0.25^{+0.01}_{-0.01}$, $\sigma=0.39^{+0.03}_{-0.03}$, $\Gamma = 1.95^{+0.20}_{-0.30}$) and the non-clustered populations ($m_c=0.30^{+0.07}_{-0.04}$, $\sigma=0.48^{+0.17}_{-0.09}$, $\Gamma = 1.73^{+0.26}_{-0.23}$). This suggests that, within the present uncertainties, the non-clustered component does not exhibit a strong mass-dependent bias relative to the clustered populations. Nevertheless, within the membership framework adopted here and the currently available kinematic information, it remains difficult to quantify the impact of field contamination, particularly in the brown dwarf regime, which could affect conclusions regarding the dynamical dispersal of substellar objects. Future spectroscopic follow-up will help address this question by providing radial velocities and independent membership confirmation, enabling a better characterisation of the kinematics and a more robust assessment of the level of contamination.

\subsection{Comparison with other IMFs in Orion}
\label{sec:allOrionIMF}

Determinations of the IMF spanning a wide stellar-mass range in Orion are 
limited to a small number of studies, each focused on specific populations. These include the Orion Nebula Cluster \citep{da_rio_initial_2012,drass_bimodal_2016}, $\sigma$ Orionis \citep{pena_ramirez_new_2012,damian2023}, 25 Orionis \citep{downes_low-mass_2014,suarez_system_2019}, and Collinder~69 in the $\lambda$ Orionis region \citep{bayo_spectroscopy_2011}. The mass ranges and IMF parameterisations reported in these studies, together with our determinations for the individual groups and the integrated OB1b IMF, are listed in \autoref{tabimfs_orionandbs4}.

\begin{figure*}
    \centering
    \begin{tabular}{cc}  \includegraphics[width=0.40\textwidth]{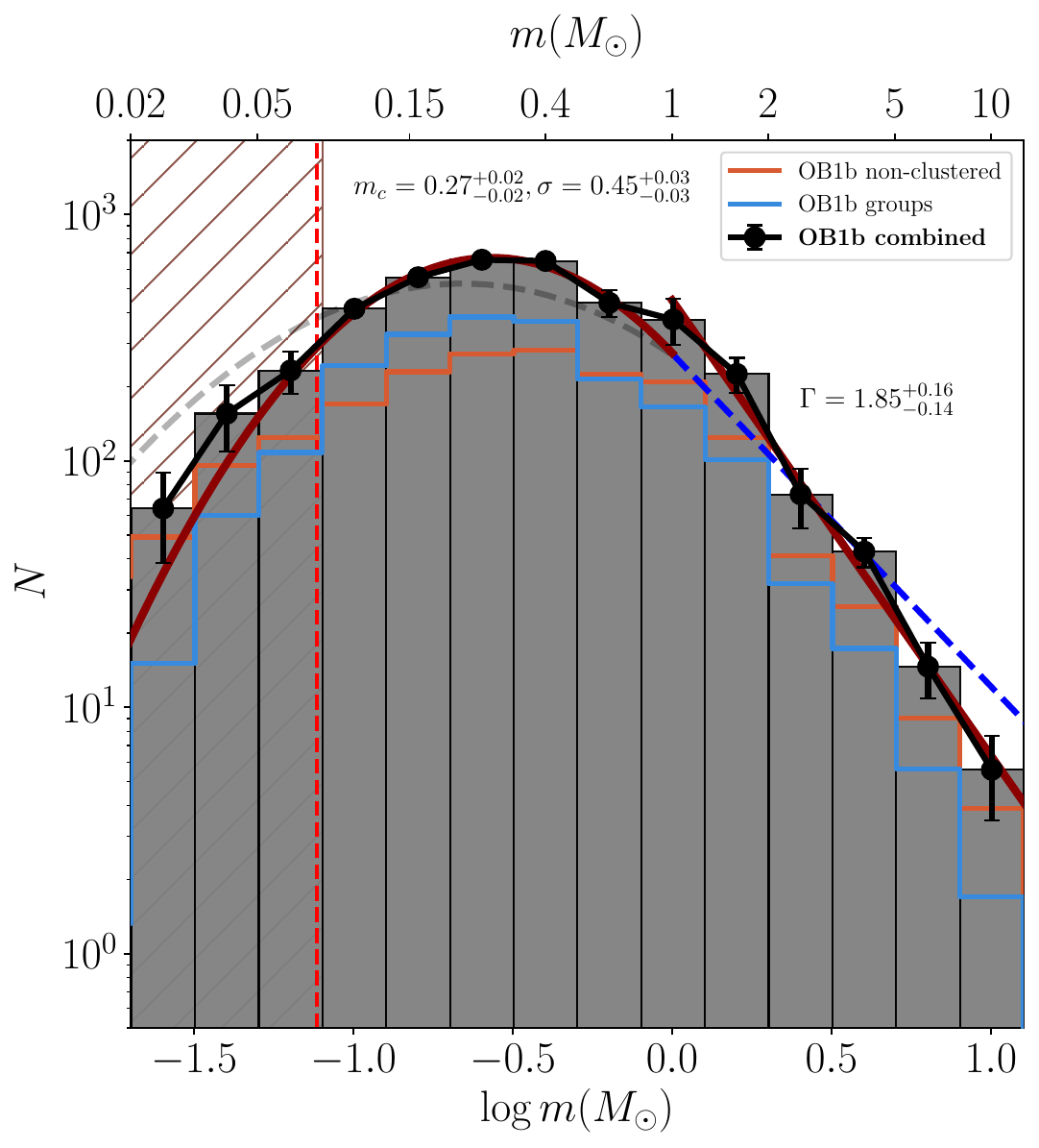} & \includegraphics[width=0.40\textwidth]{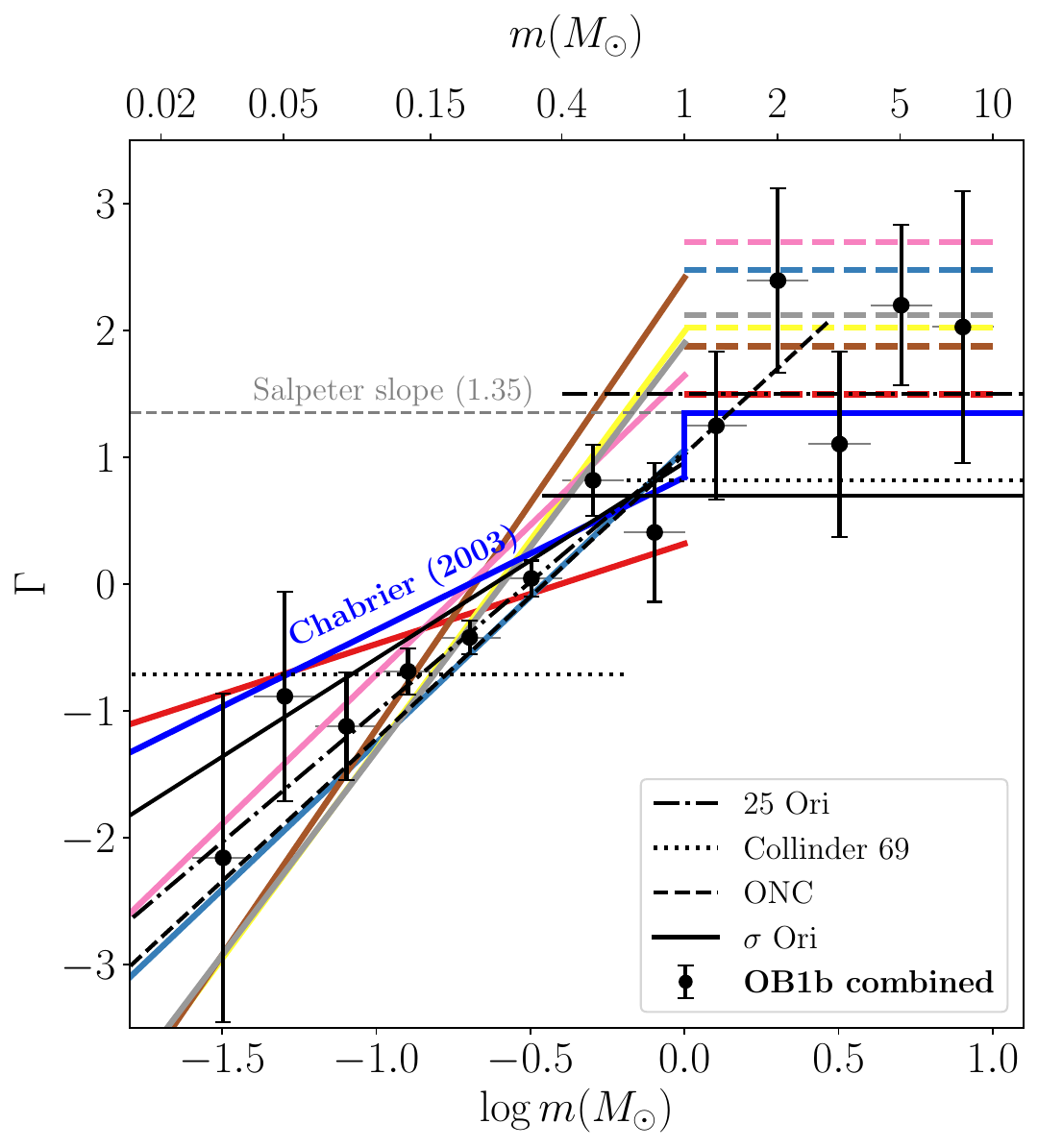}
    \end{tabular}
    \caption{Left: Combined IMF of the groups within OB1b. The IMFs of all are shown with coloured markers.     Reference IMFs and best-fit curves are as in \autoref{figimfsgroups} and the vertical red line indicates the largest mass corresponding to the validity limit magnitude among all groups ($m=0.076M_\odot$).  Right: Slope of the IMF as a function of mass for the combined IMF of the groups within OB1b compared with the canonical IMFs of \cite{chabrier_galactic_2003} and \cite{salpeter_luminosity_1955}, as well as parametrisations from other populations in Orion. References for the selected IMFs are in \autoref{tabimfs_orionandbs4}. In the case of the ONC and $\sigma$~Ori, the first IMF listed for each population in \autoref{tabimfs_orionandbs4} is shown. 
    }
    \label{fig:alphaplotandIMF}
\end{figure*}

A useful way to compare different IMFs is through the variation of their slope as a function of stellar mass \citep[e.g.][]{bastian_universal_2010,hennebelle2024}. In \autoref{fig:alphaplotandIMF} we show the slopes for the combined IMF of OB1b, the individual log-normal and power-law parametrisations for all groups and different parametrisations from the literature. For the combined IMF, we show the slopes between adjacent IMF points rather than relying on parametric forms. Although much noisier, these local slopes can reveal differences that parametrisations may hide. However, this approach requires enough objects in each IMF bin for the stochastic noise to be small, hence the use of parametrisations for the individual groups. For low masses ($m<1\,M_\odot$), the combined OB1b IMF is consistent with those of 25 Ori \citep{suarez_system_2019} and the ONC \citep{da_rio_initial_2012}, but shows steeper slopes than those reported for $\sigma$~Ori \citep{pena_ramirez_new_2012} and the Galactic field \citep{chabrier_galactic_2003}. In the high-mass regime $(m > 1\,M_\odot)$, uncertainties are larger, but consistency with 25~Ori and the Galactic field \citep{suarez_system_2019, salpeter_luminosity_1955} is clear, while tending towards steeper values than those reported for $\sigma$~Ori. We also find that the average slope of our combined IMF is consistent, within the corresponding mass ranges, with the two-slope broken power-law parametrisations of Collinder 69 reported by \cite{bayo_spectroscopy_2011}. However, this two-segment description does not fully reproduce the overall shape of our IMF across the full mass range, which exhibits a plateau near the characteristic mass.

A particularly relevant comparison is with the 25 Ori group from \cite{suarez_system_2019}, which shares several key properties with OB1b, including a similarly dispersed structure, a comparable age \citep[$\sim$7 Myr,][]{suarez_system_2019}, and a similar number of members. In addition, \cite{suarez_system_2019} adopted the same stellar evolutionary models for mass determination, thereby minimising systematic differences associated with model choice. Repeated KS tests based on random sampling from both IMFs consistently yield $p > 0.05$, and therefore do not reject the null hypothesis that both are drawn from the same parent distribution.

Overall, we find no compelling evidence that the IMF of Orion OB1b differs significantly from those of other stellar populations within the Orion complex. Although moderate differences are present, particularly in the very low-mass ($m\lesssim0.1\,M_\odot$) and high-mass ($m\gtrsim1\,M_\odot$) regimes, they remain within the level expected from present observational uncertainties. Whether these differences reflect the specific environmental conditions under which each group formed remains an open question, and addressing it would require dedicated numerical simulations as well as more spectroscopic radial velocity determinations for the objects on the low mass end of the IMF.

\section{Summary and conclusions}\label{sec:conclusions}

We present the first determination of the IMF of Orion OB1b, both for the association as a whole and for its individual stellar populations, covering masses from the brown dwarf regime ($m\sim0.076\,M_\odot$) up to massive stars ($m\sim10\,M_\odot$), using a probability-weighted framework based on Gaia DR3 photometry and astrometry. Our main conclusions are summarised as follows:

\begin{itemize}

    \item  By applying a clustering algorithm to Gaia DR3 astrometry, 
    we identified eight kinematically distinct groups within
    Orion OB1b.
    These results reinforce previous evidence that Orion OB1b is composed of multiple overlapping stellar populations 
    \citep[e.g.][]{kounkel_apogee-2_2018,briceno_cida_2019,zari2019,chen2020,swiggum2021, sanchez2024}. 
    We established the correspondence between our groups and the previously 
    reported substructures in an attempt to provide a more unified view of 
    OB1b (Section \ref{sec:memberships}).
    
    \item By analysing the association of spectroscopically confirmed young 
    stars from \cite{briceno_cida_2019} with the identified groups, 
    we found evidence for a population of sources not associated with any 
    group. We interpret this as a non-clustered component within OB1b, although 
    the current data do not allow us to fully distinguish between a genuinely 
    dispersed young population and sources not associated with more compact groups 
    by the adopted clustering procedure (\autoref{figclustered_candidates}).

    \item We developed a probabilistic framework to account for the strong overlap 
    between populations in the region by modelling the groups, the non-clustered 
    component and the Galactic field by using mixtures of multivariate Gaussian 
    distributions in the astrometric space. This allowed us to assign 
    individual membership probabilities to all sources in the science sample 
    to each component of OB1b (\autoref{tab:numbersources}).
    
    \item From isochrone fitting and using the spatial distribution of spectroscopically 
    determined extinctions, we derived ages for the different components, ranging 
    from $\sim5$ Myr to $\sim15$~Myr. The inferred age sequence is consistent 
    with the relative fraction of sources showing near-infrared excesses 
    within each group, which follows an exponential decay consistent with previous 
    results for Orion OB1b from \cite{briceno_cida_2019} (\autoref{figinfrared_excesses}).
    
    \item Adopting the ages derived for each component and using 
    pre-main-sequence evolutionary models to convert magnitudes and 
    colours into masses, we computed the probability-weighted IMFs 
    of all groups and the non-clustered component (\autoref{figimfsgroups}), 
    reaching the brown dwarf regime in all cases, although incompleteness remains 
    at the lowest masses. The resulting IMFs are broadly consistent with canonical 
    parameterisations and previous determinations across Orion (\autoref{fig:alphaplotandIMF}). 
    Although KS tests suggest statistical differences among some components, 
    we cannot yet rule out methodological and observational effects related 
    to membership assignment, particularly in the sparser components.

    \item From a virial analysis accounting for observational uncertainties in the velocity dispersion, we conclude that all groups are gravitationally unbound. This result motivates further investigation of the dynamical evolution of OB1b to determine whether these groups formed as bound systems that subsequently dispersed or were born unbound. Such processes may be relevant for understanding the origin of the non-clustered population and the observed differences between clustered and non-clustered components.
    
\end{itemize}

The presented analysis illustrates the value of probabilistic membership assignment for disentangling overlapping stellar populations and for deriving robust IMFs in complex star-forming environments. The use of spectroscopically confirmed members proved essential as a control sample for constraining ages, extinction, and the physical interpretation of populations identified from photometric and astrometric surveys affected by contamination from older field stars. Future data releases from Gaia, together with LSST and dedicated spectroscopic follow-up, will extend this approach towards lower masses and improved kinematic constraints, enabling more complete determinations of the IMF across a broader range of young stellar populations.

\begin{acknowledgements}
RC gratefully acknowledges the scholarship Becas de apoyo a docentes para estudios de posgrado from the Comisión Académica de Posgrado, Universidad de la República, Uruguay, for financial support that made this work possible. RC and JJD also acknowledge the Programa de Desarrollo de Ciencias Básicas (PEDECIBA). We thank Eleonora Zari for generously sharing part of the data used in this work, and Cecilia Mateu, Mauro Cabrera, Bruno Domínguez, and Selena Seidel for valuable discussions throughout its development. Finally, we thank the anonymous referee for providing useful comments that have strengthened the quality and presentation of this work.

\end{acknowledgements}

\bibliographystyle{aa}
\bibliography{references}

\begin{appendix}

\onecolumn

\section{The covariance matrix}\label{app:matrix}

\begin{align}
    \resizebox{.5\hsize}{!}{$X_{err}(\mathbf{u})=\left(\begin{array}{lllll}
    \sigma_{\alpha}^2&\sigma_\alpha\sigma_\delta\rho_{\alpha\delta}&\sigma_{\alpha}\sigma_{\mu_{\alpha*}}\rho_{\alpha\mu_{\alpha*}}&\sigma_{\alpha}\sigma_{\mu_\delta}\rho_{\alpha\mu_\delta}&\sigma_\alpha\sigma_\varpi\rho_{\alpha\varpi} \\
    \sigma_\delta\sigma_\alpha\rho_{\alpha\delta}&\sigma_\delta^2&\sigma_\delta\sigma_{\mu_\alpha*}\rho_{\delta\mu_{\alpha*}}&\sigma_\delta\sigma_{\mu_\delta}\rho_{\delta\mu_\delta}&\sigma_\delta\sigma_\varpi\rho_{\delta\varpi} \\
\sigma_{\mu_{\alpha*}}\sigma_{\alpha}\rho_{\alpha\mu_{\alpha*}}&\sigma_{\mu_{\alpha*}}\sigma_{\delta}\rho_{\delta\mu_{\alpha*}}&\sigma_{\mu_{\alpha *}}^2 & \sigma_{\mu_{\alpha *}} \sigma_{\mu_\delta} \rho_{\mu_\alpha \mu_\delta} & \sigma_{\mu_{\alpha *}} \sigma_{\varpi} \rho_{\varpi \mu_{\alpha *}} \\
\sigma_{\mu_\delta}\sigma_{\alpha}\rho_{\alpha\mu_\delta}&\sigma_{\mu_{\delta}}\sigma_\delta\rho_{\delta\mu_\delta}&\sigma_{\mu_\delta} \sigma_{\mu_{\alpha *}} \rho_{\mu_\alpha \mu_\delta} & \sigma_{\mu_\delta}^2 & \sigma_{\mu_\delta} \sigma_{\varpi} \rho_{\varpi \mu_\delta} \\
\sigma_{\varpi}\sigma_\alpha\rho_{\alpha\varpi}&\sigma_{\varpi}\sigma_{\delta}\rho_{\delta\varpi}&\sigma_{\varpi} \sigma_{\mu_{\alpha *}} \rho_{\varpi \mu_{\alpha *}} & \sigma_{\varpi} \sigma_{\mu_\delta} \rho_{\varpi \mu_\delta} & \sigma_{\varpi}^2 .
\end{array}\right)$}
\end{align}

\section{Comparisons with the literature}\label{app:comparLit}

\begin{figure*}[h!]
    \sidecaption
    \includegraphics[width=12cm]{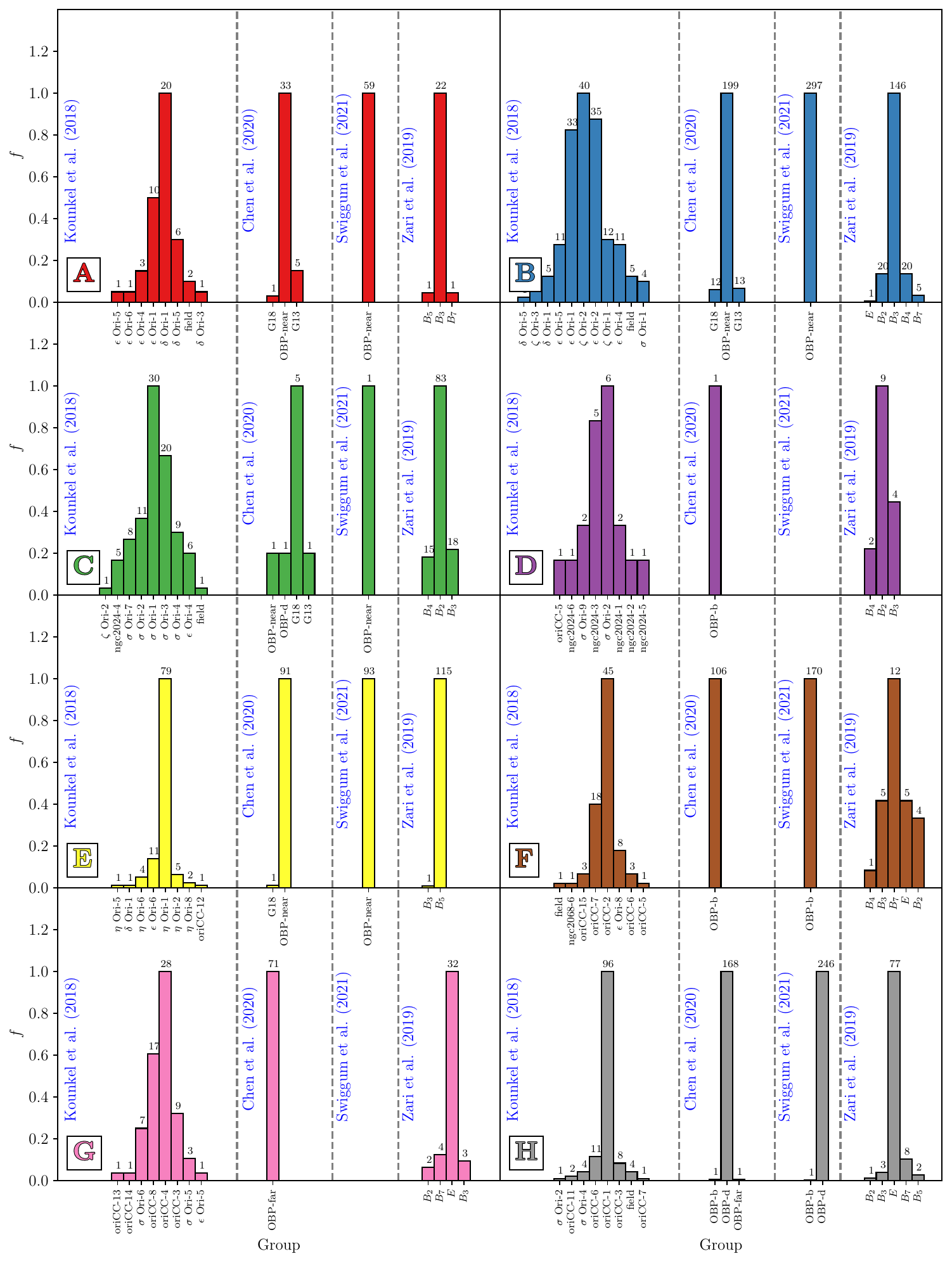}
    \caption{Distribution of labels assigned to members of the clustered sample by different works from the literature. The colours and plot labels correspond to the groups identified in this work, following the same colour scheme as \autoref{figclustered_candidates}. The normalised histograms indicate how many sources within each group are assigned to the different groups shown below each bin, coming from the references shown in blue.}
    \label{fig:literaturecomparisons}
\end{figure*}

\begin{figure*}[h]
    \sidecaption
    \includegraphics[width=12cm]{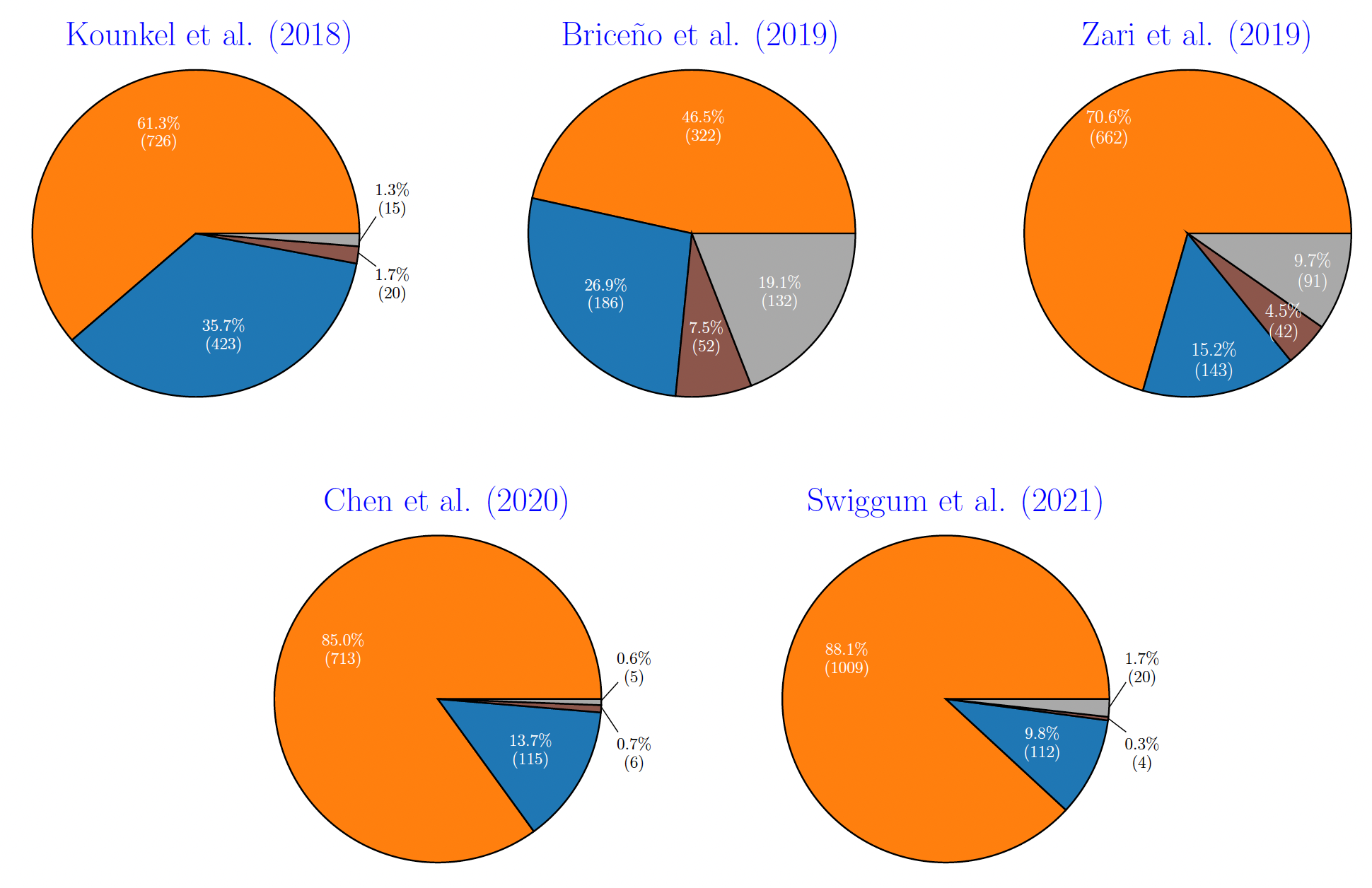}
    \caption{Fraction of sources from different samples from the literature (indicated above each pie chart) classified as members of any of the groups found in this work (orange), the non-clustered component (blue), the field component (brown), or not recovered in our catalogue (grey). A source is assigned to a given component if its membership probability for that component exceeds that of all others.}
    \label{fig:piecharts}
\end{figure*}

\section{Extinction coefficients}\label{app:ExtCoef}

\begin{table}[H]
\centering
\caption{Coefficients of the fit given by \autoref{eqexcfit}.}
\begin{adjustbox}{width=0.95\textwidth}\begin{tabular}{ccccccccccccccccc}
\hline\hline
$k_{band}$ & $x$ & $c_{00}$ & $c_{10}$ & $c_{20}$ & $c_{30}$ & $c_{40}$ & $c_{01}$ & $c_{02}$ & $c_{03}$ & $c_{04}$ & $c_{11}$ & $c_{12}$ & $c_{13}$ & $c_{21}$ & $c_{22}$ & $c_{31}$ \\
\hline
$k_G$ & $BP-RP$ & 0.9888 & -0.1449 & 0.0337 & 0.0023 & -0.0096 & 0.0030 & 0.0060 & -0.0004 & -0.0012 & 0.0001 & -0.0009 & 0.0005 & -0.0000 & 0.0000 & 0.0000 \\
$k_{BP}$ & $BP-RP$ & 1.1373 & -0.1056 & 0.0284 & 0.0105 & -0.0142 & 0.0038 & 0.0087 & -0.0060 & 0.0015 & -0.0003 & -0.0018 & 0.0021 & -0.0010 & 0.0003 & -0.0000 \\
$k_{RP}$ & $BP-RP$ & 0.6793 & -0.0191 & 0.0025 & -0.0050 & 0.0025 & -0.0003 & -0.0009 & 0.0020 & -0.0012 & 0.0002 & 0.0004 & -0.0007 & 0.0005 & -0.0001 & 0.0000 \\
$k_G$ & $G-RP$ & 0.9847 & -0.2169 & 0.0414 & -0.0179 & -0.0182 & -0.0007 & -0.0170 & -0.0221 & 0.0040 & 0.0006 & 0.0062 & 0.0130 & -0.0027 & -0.0003 & -0.0000 \\
$k_{BP}$ & $G-RP$ & 1.1326 & -0.1527 & 0.0370 & 0.0158 & -0.0263 & -0.0015 & -0.0112 & -0.0369 & 0.0070 & 0.0009 & 0.0229 & 0.0203 & -0.0037 & -0.0005 & -0.0000 \\
$k_{RP}$ & $G-RP$ & 0.6797 & -0.0184 & -0.0013 & -0.0236 & -0.0004 & 0.0019 & -0.0376 & 0.0371 & -0.0082 & 0.0002 & 0.0156 & -0.0248 & 0.0056 & -0.0004 & 0.0000 \\
\hline\hline
\end{tabular} \end{adjustbox}

   \tablefoot{In all cases, colours employed in the fit are reddened colours.} \label{tab_extinction_coefficients}

\end{table}

\clearpage

\section{Colour-magnitude diagram of each group}

\begin{figure*}[h!]
    \sidecaption
    \includegraphics[width=12cm]{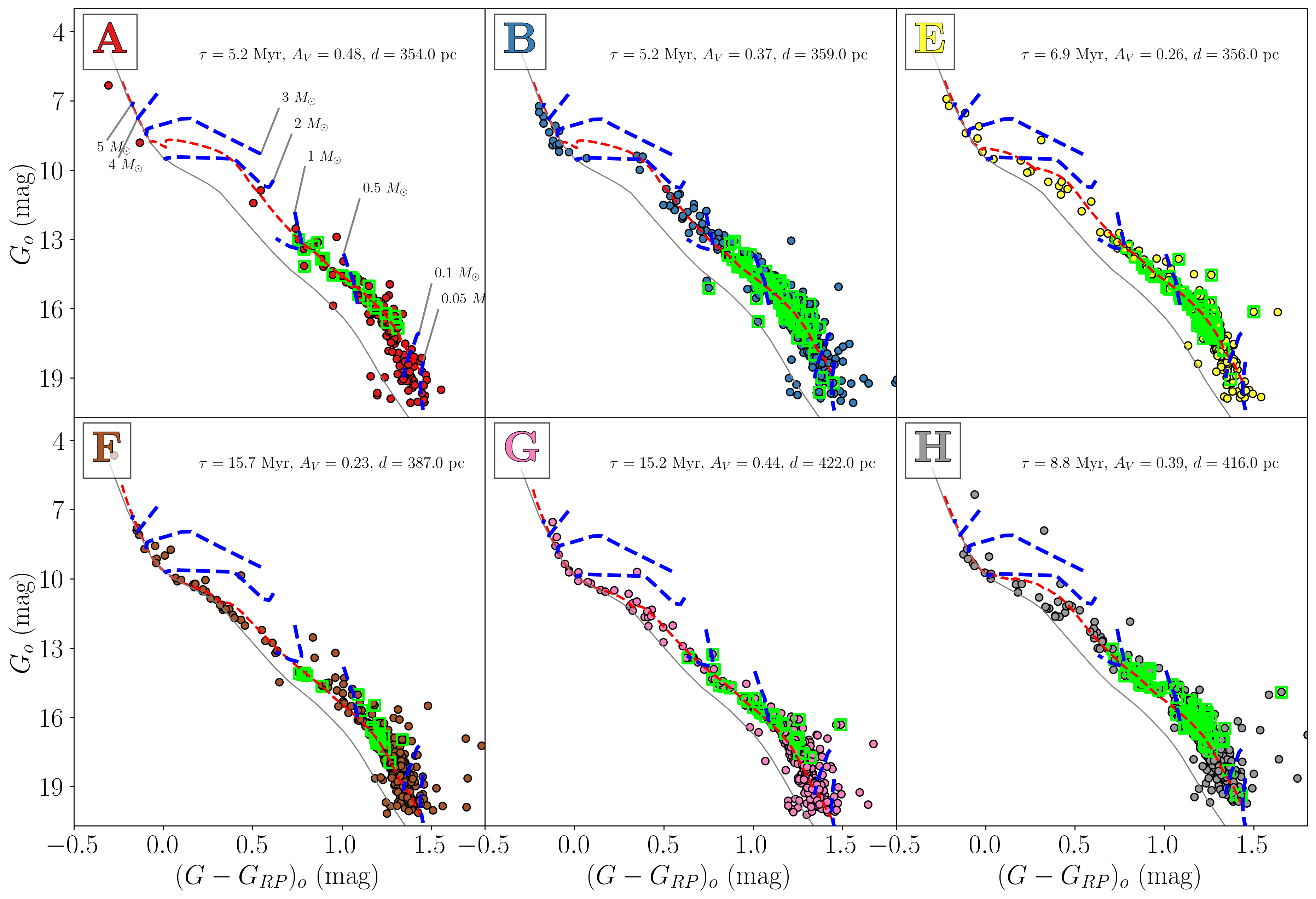}
    \caption{CMD of most probable candidates of each group (coloured dots) and the spectroscopic sample (green empty squares). The best-fitting isochrone is shown with a red dashed line and the solar-metallicity ZAMS corresponding to the median distance of group candidates of each group is shown in grey. In blue dashed lines are shown, from top to bottom, the evolutionary tracks from 1 to 20 Myr of stars with 5, 4, 3, 2, 1, 0.5, 0.1, and 0.05 $M_\odot$. On top of the plot, the fitted age $\tau$, the median $A_V$, and the median distance $d$ of the group candidates are shown.}
    \label{figCMDallgroups}
\end{figure*}

\section{Colour-colour diagrams of each group}

\begin{figure*}[h!]
    \sidecaption
    \includegraphics[width=12cm]{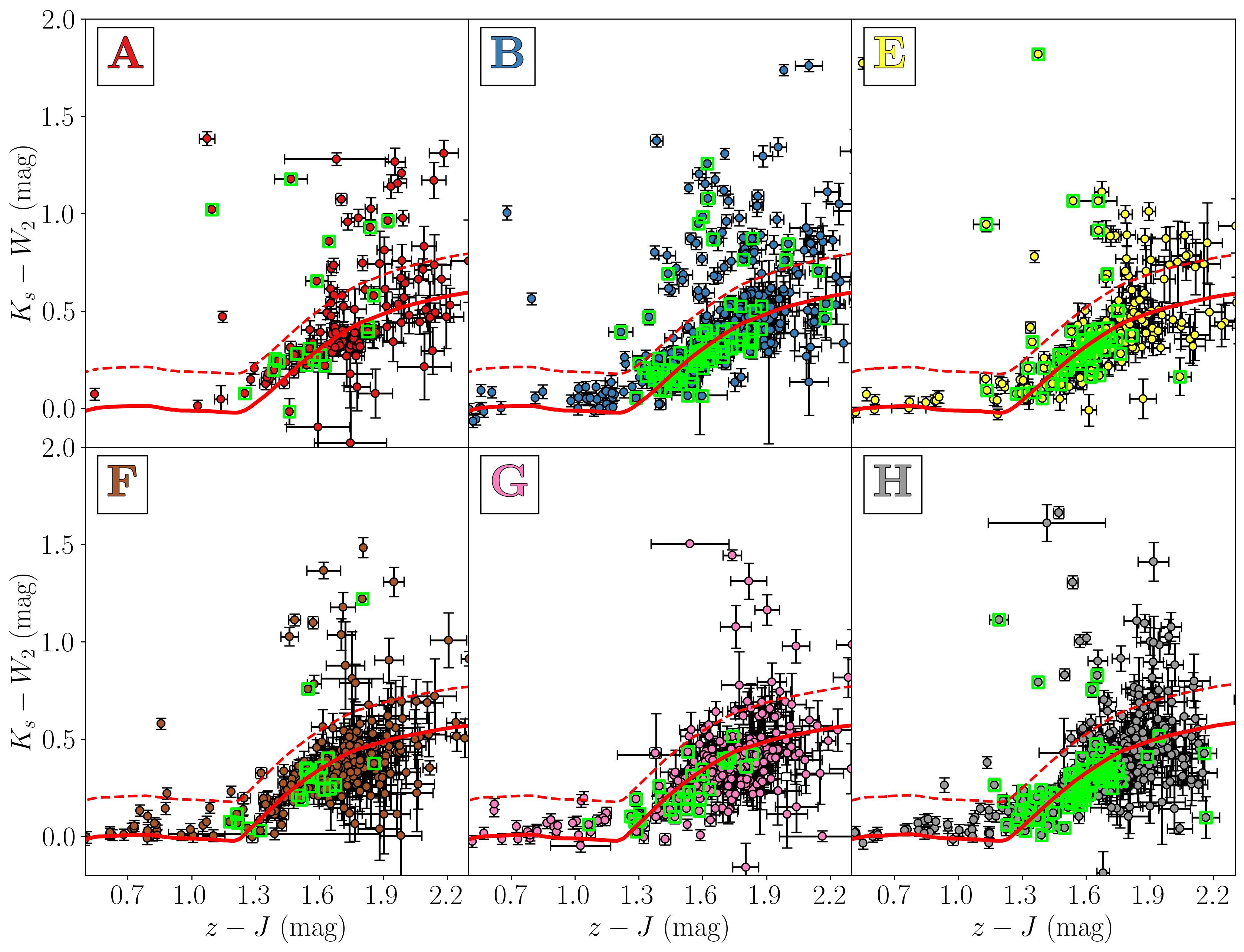}
    \caption{Colour-colour diagrams $W_1-W_2$ vs $z-J$ of most probable candidates of each group (coloured dots). Green squares: the spectroscopic sample. Red solid line: isochrone corresponding to the age of each group. Red dashed line: $0.1~$mag offset in $W_1-W_2$.}
    \label{figCCallclusters}
\end{figure*}

\clearpage

\section{Luminosity functions}

\begin{figure*}[!h]
    \sidecaption
    \includegraphics[width=12cm]{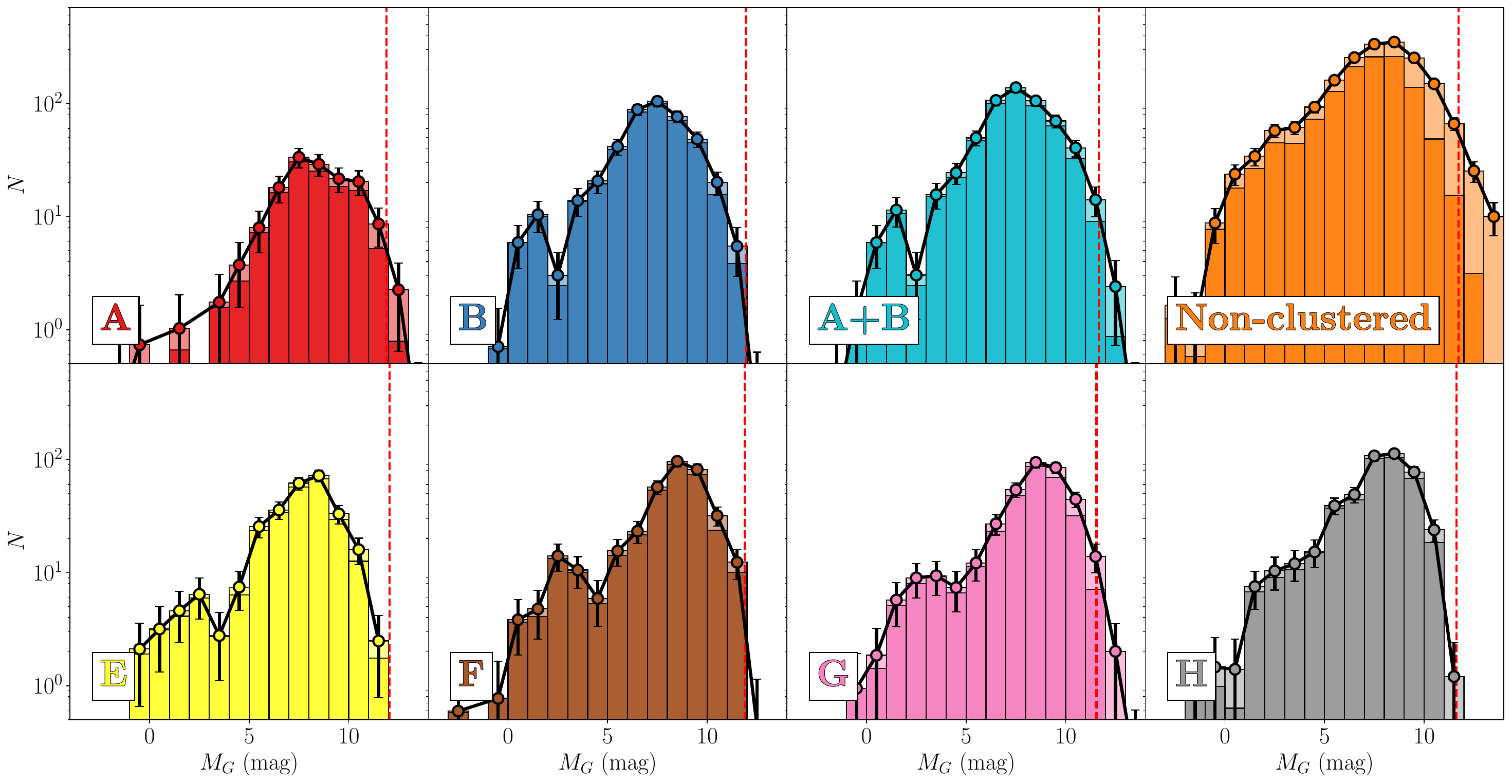}
    \caption{Probability-weighted LFs of all groups identified in Orion OB1b. Light colors: complete science sample. Dark colors: most probable candidates. Red vertical lines: absolute magnitude corresponding to the approximate validity limit of $G=20$~mag (Section \ref{sec:indmemberships}).}
    \label{figlfsgroups}
\end{figure*}

\section{Used parametrisations of the IMF}

\begin{enumerate}[$(i)$]
    \item Power-law distribution for $m>1M_\odot$ \citep[e.g.,][]{salpeter_luminosity_1955,chabrier_galactic_2003}:
    \begin{equation}
        \xi{(\log{m})} \propto m^{-\Gamma}
    \end{equation}
    where $\Gamma$ is the power-law slope.  \\
    \item Log-normal distribution for $m<1~M_\odot$ \citep{miller_initial_1979,chabrier_galactic_2003}:
    \begin{equation}
        \xi{(\log{m})}\propto e^{\frac{(\log{m}-\log{m_c})^2}{2\sigma^2}}
    \end{equation}
    where $m_c$ is the characteristic mass and $\sigma$ is the standard deviation. \\
    \item Tapered power-law over the whole valid mass range \citep{de_marchi_stellar_2005}:
    \begin{equation}
        \xi{(\log{m})} \propto m^{-\Gamma}\left[1-e^{-(m/m_p)^\beta}\right]
    \end{equation}
    where $m_p$ is the peak mass, $\Gamma$ the power-law index and $\beta$ the tapering exponent.
\end{enumerate} 

\label{sec:parametrisations}

\clearpage

\section{IMF parametrisations within Orion}

\begin{table*}[h!]
    \centering
    \caption{Different IMFs parametrisations in the Orion complex, including those of Orion OB1b obtained in this work.}
    \resizebox{\textwidth}{!}{
    \begin{tabular}{ccccccccccc}
    \hline\hline
    Population & \multicolumn{3}{c}{Lognormal} & \multicolumn{2}{c}{Power-law} & \multicolumn{4}{c}{Tapered power-law} & Ref\\
    \hline
    & $m_c$ & $\sigma$ & $m$ range $(M_\odot)$ & $\Gamma$ & $m$ range $(M_\odot)$ & $\Gamma$ & $m_p$ & $\beta$ & $m$ range $(M_\odot)$ & \\
    & & & & & & & & & & \\
    \multirow{3}{*}{$\sigma$ Ori} & $0.24\pm0.09$ & $0.53\pm0.19$ & 0.006-1 & $0.70\pm0.20$ & 0.35-19 & & & & & \multirow{2}{*}{1}\\
    & $0.27\pm0.09$ & $0.63\pm0.15$ & 0.006-19 & & & & & & & \\
    & $0.20\pm0.06$ & $0.53\pm0.06$ & 0.004-18.7 & $0.81\pm0.18$ & 0.19-18.7 & & & & & 2 \\
    & & & & & & & & & & \\
    Collinder 69 & & & & $0.82\pm0.05$ & 0.65-25 & & & & & 3\\
    & & & & & & & & & & \\
    \multirow{2}{*}{ONC} & $0.35\pm0.02$ & $0.44\pm0.05$ & $0.025-3$ & $0.60\pm0.33$ & 0.30-3 & & & & & \multirow{2}{*}{4}\\
    & $0.28\pm0.02$ & $0.38\pm0.01$ & & $1.30\pm0.09$ & 0.17-3 & & & & & \\
    & & & & & & & & & & \\
    25 Ori & $0.31\pm0.04$ & $0.46\pm0.05$ & 0.012-1 & $1.50\pm0.11$ & 0.4-13.1 & $1.10\pm0.09$ & $0.31\pm0.03$ & $2.11\pm0.09$ & 0.012-13.1 & 5\\
    & & & & & & & & & & \\
    OB1b $A$ & $0.34^{+0.1}_{-0.06}$ & $0.74^{+0.35}_{-0.17}$ & 0.033-1 & $1.35^{+0.55}_{-0.4}$ & 1-[8-12.7] & $1.59^{+0.27}_{-0.31}$ & $0.49^{+0.13}_{-0.10}$ & $2.34^{+0.43}_{-0.35}$ & 0.033-[8-12.7] & 6\\
    & & & & & & & & & & \\
    OB1b $B$ & $0.35^{+0.02}_{-0.02}$ & $0.43^{+0.03}_{-0.03}$ & 0.032-1 & $1.64^{+0.41}_{-0.34}$ & 1-[8-12.7] & $1.40^{+0.11}_{-0.10}$ & $0.37^{+0.02}_{-0.02}$ & $2.46^{+0.12}_{-0.11}$ & 0.032-[8-12.7] & 6\\
    & & & & & & & & & & \\
    OB1b $A+B$ & $0.35^{+0.03}_{-0.02}$ & $0.48^{+0.04}_{-0.03}$ & 0.033-1 & $1.8^{+0.37}_{-0.32}$ & 1-[8-12.7] & $1.48^{+0.10}_{-0.09}$ & $0.39^{+0.03}_{-0.03}$ & $2.41^{+0.10}_{-0.10}$ & 0.033-[8-12.7] & 6\\
    & & & & & & & & & & \\
    OB1b $E$ & $0.25^{+0.01}_{-0.01}$ & $0.36^{+0.02}_{-0.02}$ & 0.035-1 & $1.63^{+0.41}_{-0.35}$ & 1-[5-8] & $1.05^{+0.10}_{-0.09}$ & $0.23^{+0.03}_{-0.03}$ & $2.37^{+0.13}_{-0.12}$ & 0.035-[5-8] & 6\\
    & & & & & & & & & & \\
    OB1b $F$ & $0.21^{+0.01}_{-0.01}$ & $0.35^{+0.02}_{-0.02}$ & 0.065-1 & $1.68^{+0.44}_{-0.3}$ & 1-[8-12.7] & $0.92^{+0.06}_{-0.06}$ & $0.17^{+0.01}_{-0.01}$ & $2.24^{+0.20}_{-0.19}$ & 0.065-[8-12.7] & 6\\
    & & & & & & & & & & \\
    OB1b $G$ & $0.20^{+0.01}_{-0.01}$ & $0.43^{+0.05}_{-0.04}$ & 0.076-1 & $1.88^{+0.43}_{-0.35}$ & 1-[5-8] & $1.06^{+0.10}_{-0.08}$ & $0.22^{+0.05}_{-0.03}$ & $1.81^{+0.18}_{-0.16}$ & 0.076-[5-8] & 6\\
    & & & & & & & & & & \\
    OB1b $H$ & $0.26^{+0.01}_{-0.01}$ & $0.37^{+0.02}_{-0.02}$ & 0.055-1 & $1.92^{+0.38}_{-0.29}$ & 1-[8-12.7] & $1.09^{+0.07}_{-0.06}$ & $0.26^{+0.02}_{-0.02}$ & $2.19^{+0.08}_{-0.08}$ & 0.055-[8-12.7] & 6\\
    & & & & & & & & & & \\
    OB1b groups & $0.25^{+0.01}_{-0.01}$ & $0.39^{+0.03}_{-0.03}$ & 0.076-1 & $1.95^{+0.20}_{-0.30}$ & 1-10 & $1.16^{+0.09}_{-0.08}$ & $0.28^{+0.04}_{-0.03}$ & $1.98^{+0.13}_{-0.11}$ & 0.076-[8-12.7] & 6\\
    & & & & & & & & & & \\
    OB1b non-clustered & $0.30^{+0.07}_{-0.04}$ & $0.48^{+0.17}_{-0.09}$ & 0.076-1 & $1.73^{+0.26}_{-0.23}$ & 1-[8-12.7] & $1.16^{+0.09}_{-0.08}$ & $0.42^{+0.23}_{-0.15}$ & $1.98^{+0.13}_{-0.11}$ & 0.076-[8-12.7] & 6\\
     & & & & & & & & & & \\
    OB1b combined & $0.27^{+0.02}_{-0.02}$ & $0.45^{+0.03}_{-0.03}$ & 0.076-1 & $1.85^{+0.16}_{-0.14}$ & 1-[8-12.7] & $1.13^{+0.23}_{-0.14}$ & $0.36^{+0.15}_{-0.08}$ & $1.83^{+0.16}_{-0.26}$ & 0.076-[8-12.7] & 6\\
    \hline\hline
    \end{tabular}}
    \label{tabimfs_orionandbs4}

    \tablefoot{
    For the IMFs from this work, the lower mass limits correspond to the completeness mass associated with the average distances, ages and extinctions of the corresponding group, and the upper mass limits correspond to the mass interval of the most massive bin with at least one detection.
    }

    \tablebib{(1)~\citet{pena_ramirez_new_2012}; (2)~\citet{damian2023}; (3)~\citet{bayo_spectroscopy_2011}; (4)~\citet{da_rio_initial_2012}; (5)~\citet{suarez_system_2019}; (6)~ This work
    
    }
\end{table*}

\clearpage

\section{Comparisons between OB1b IMFs}

\begin{figure}[h!]
    \sidecaption
    \includegraphics[width=12cm]{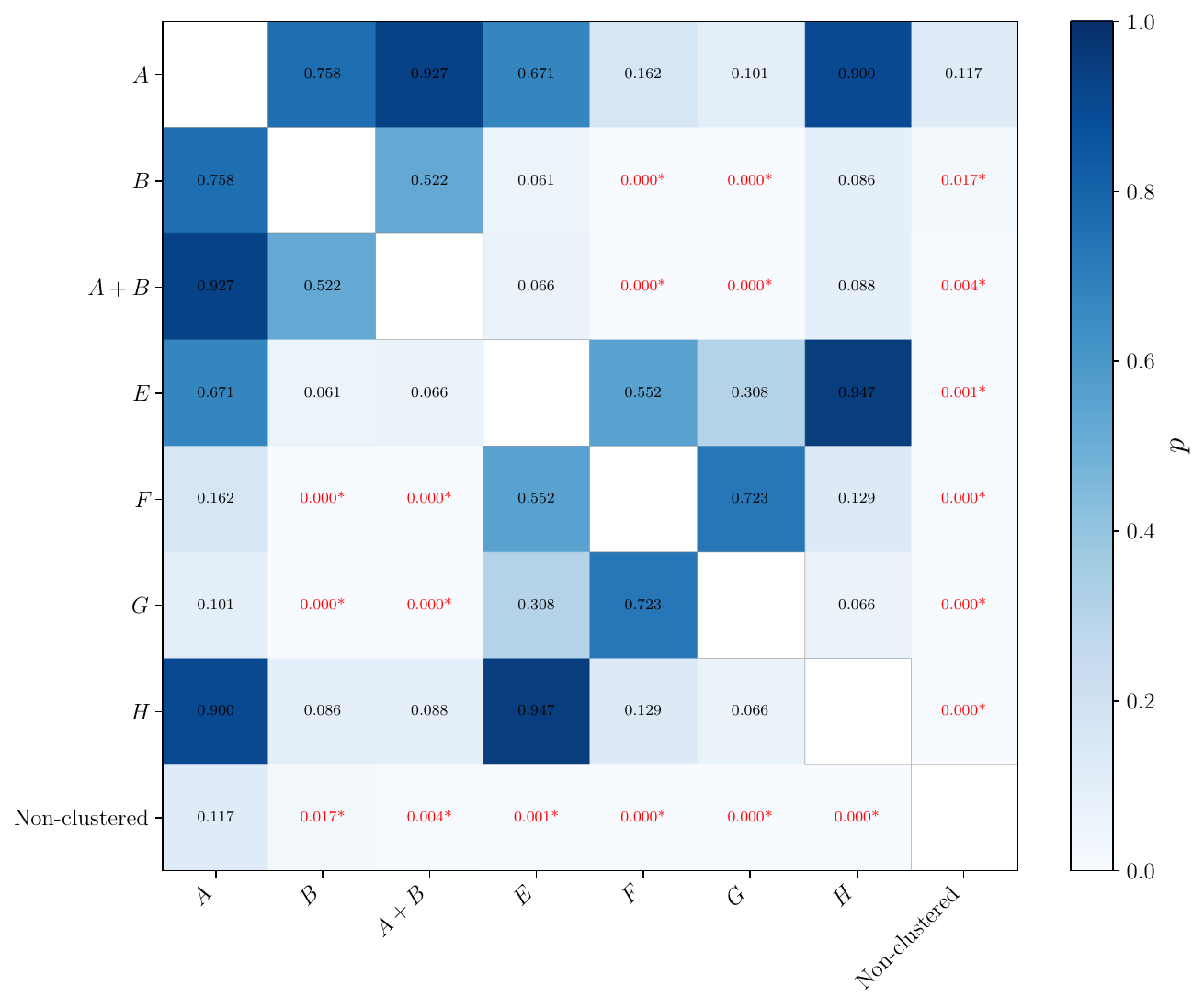}
    \caption{Confusion matrix with the $p$-values resulting from K-S tests between the obtained IMFs in this work.}
    \label{figkstest}
\end{figure}

\end{appendix}

\end{document}